\documentclass[12pt,preprint]{aastex}

\defcitealias{bureau-etal99}{BFPM99}
\defcitealias{pfitzner99}{P99}
\defcitealias{dubinski92}{D92}
\defcitealias{mb03}{MB03}
\newcommand{\kmskpc}{\ensuremath{\mathrm{km~s^{-1}~kpc^{-1}}}}
\newcommand{\hkmskpc}{\ensuremath{h~\mathrm{km~s^{-1}~kpc^{-1}}}}
\newcommand{\lcdm}{{$\mathrm{\Lambda}$CDM}}
\newcommand{\rhocrit}{\ensuremath{\rho_{\mathrm{crit}}}}
\newcommand{\ntnif}{{\objectname{NGC~2915}}}
\newcommand{\rvir}{\ensuremath{r_{\mathrm{vir}}}}
\newcommand{\hmyr}{\ensuremath{h^{-1}~\mathrm{Myr}}}

\newcommand{\thboot}{\ensuremath{\theta_{\mathrm{boot}}}}
\newcommand{\therrN}{\ensuremath{\theta_{\mathrm{err},N}}}
\newcommand{\therrba}{\ensuremath{\theta_{\mathrm{err},b/a}}}
\newcommand{\therr}{\ensuremath{\theta_{\mathrm{err}}}}
\newcommand{\hmsun}{\ensuremath{h^{-1}~\mathrm{M_{\Sun}}}}
                                                                                
\shorttitle{Figure Rotation of Dark Matter Halos}
\shortauthors{Bailin and Steinmetz}

\begin{document}

\title{Figure Rotation of Cosmological Dark Matter Halos}
\author{Jeremy Bailin\altaffilmark{1} and 
Matthias Steinmetz\altaffilmark{2,3}}
\affil{Steward Observatory, University of Arizona, 933 N Cherry Ave,
 Tucson, AZ 85721 USA\\and\\
Astrophysikalisches Institut Potsdam, An der Sternwarte 16,
 D-14482 Potsdam, Germany}
\altaffiltext{1}{jbailin@as.arizona.edu}
\altaffiltext{2}{msteinmetz@aip.de}
\altaffiltext{3}{David and Lucile Packard Fellow}

\begin{abstract}
We have analyzed galaxy and group-sized dark matter halos formed in a
high resolution \lcdm\ numerical N-body simulation
in order to study the rotation of the triaxial figure,
a property in principle independent of the angular momentum of the
particles themselves.
Such figure rotation may have observational consequences, such
as triggering spiral structure in extended gas disks.
The orientation of the major axis is compared at 5 late snapshots
of the simulation.
Halos with significant substructure or that appear otherwise
disturbed are excluded from the sample.
We detect smooth figure rotation in 278 of the 317 halos in
the sample. The pattern speeds follow a log normal distribution
centred at $\Omega_p = 0.148~\hkmskpc$ with a width of 0.83.
These speeds are an order of magnitude
smaller than required to explain the spiral
structure of galaxies such as \ntnif.
The axis about which the figure rotates aligns very well with
the halo minor axis, and also reasonably well with its
angular momentum vector.
The pattern speed is correlated with the halo spin parameter $\lambda$,
but shows no correlation with the halo mass.
The halos with the highest pattern speeds show particularly strong alignment
between their angular momentum vectors and their
figure rotation axes.
The figure rotation is coherent outside 0.12~\rvir. The measured
pattern speed and degree of internal alignment of the figure rotation
axis drops in the innermost region of the halo, which may be 
an artifact of the numerical force softening.
The axis ratios show a weak tendency to become more spherical with time.
\end{abstract}


\section{Introduction}

Although there have been many theoretical studies of the shapes of cosmological
dark matter halos \citep[e.g.][]{dc91,warren-etal92,cl96,js02},
there has been relatively little
work done on how those figure shapes evolve with time, in particular,
whether the orientation of a triaxial
halo stays fixed, or whether the figure rotates.
While the orientation of the halo
can clearly change during a major merger, it is
not known whether the orientation changes
in between cataclysmic events. Absent any theoretical prediction
one way or the other, it is usually assumed that the figure orientation 
 of triaxial halos  remain fixed when in isolation
\citep[e.g.][]{subramanian88,johnston-etal99,ls03}

Early work at detecting figure rotation
in simulated halos was done by \citet{dubinski92}
(hereafter \citetalias{dubinski92}).
While examining the effect of tidal shear on halo shapes,
he found that in all 14 of his 1--$2 \times 10^{12}~\mathrm{M_{\Sun}}$
halos, the
direction of the major axis rotated uniformly around the minor axis.
The period of rotation varied from halo to halo, and ranged from
4~Gyr at the fast end to 50~Gyr at the slow end, or equivalently
the angular velocities ranged between
0.1~and 1.6~\kmskpc. \footnote{It may be more intuitive to think
of angular velocity in units of radians~$\mathrm{Gyr^{-1}}$
rather than the common unit of pattern speed, \kmskpc.
Fortunately, the two units give almost identical numerical values.}
It is difficult to draw statistics from this small sample size,
especially since the initial conditions of this simulation were
not drawn from cosmological models, but were performed in a small isolated box
with the linear tidal field of the external matter
prescriptively superimposed \citep{dc91}.
Given that the main result of \citetalias{dubinski92} is that the tidal shear
may have a significant impact on the shapes of halos, it is clearly
important to do such studies using self-consistent
cosmological initial conditions.

Recent studies of figure rotation come from \citet{bureau-etal99}
(\citetalias{bureau-etal99}) and \citet{pfitzner99}
(\citetalias{pfitzner99}).
\citetalias{pfitzner99} compared the orientation of cluster mass
halos in two snapshots spaced 500~Myr apart in an SCDM simulation
($\Omega=1$, $\Lambda=0$, $h=0.5$).
He detected rotation of the major axis in
$\sim 5\%$ of them, and argued
that the true fraction with figure rotation is probably higher.
\citetalias{bureau-etal99} presented one of these halos
which was extracted from its cosmological
surroundings and left to evolve in isolation for 5~Gyr.
During that time, the major axis rotated around the
minor axis uniformly at all radii
at a rate of 60\degr\ $\mathrm{Gyr^{-1}}$, or about 1~\kmskpc.

There may be observational consequences to a dark matter halo whose figure
rotates. \citetalias{bureau-etal99} suggested that triaxial figure rotation
is responsible for the spiral structure
of the blue compact dwarf galaxy \ntnif. Outside of the
optical radius, \ntnif\ has a large \ion{H}{1}\ disk
extending to over 22 optical disk scale lengths \citep{meurer-etal96}.
The gas disk shows clear evidence of a bar, and a spiral pattern extending
over the entire radial extent of the disk. \citetalias{bureau-etal99} argue
that the observed gas surface density is too low
for a bar or spiral structure to form by gravitational instability,
and that there is no evidence
of an interacting companion to trigger the pattern.
They propose that the pattern may instead be triggered by a rotating
triaxial halo.

\citet{bf02} followed this up with Smoothed Particle Hydrodynamics (SPH)
simulations of a disk inside a halo whose figure rotates, and showed that a
triaxial halo with a flattening of $b/a=0.8$ and a pattern
speed of $3.84~\kmskpc$ could trigger spiral patterns in the disk.
\citet{mb03} (hereafter \citetalias{mb03}) found that
in detail, the observations of \ntnif\ are better fit by increasing
the disk mass by an order of magnitude
\citep*[for example, if most of the hydrogen is molecular, e.g.][]{pcm94},
but that a triaxial halo with $b/a\approx 0.85$ and a pattern speed
of between 6.5~and $8.0~\kmskpc$ also provides an acceptable fit.

\citetalias{mb03} concluded that if the halo were undergoing solid body
rotation at this rate, its spin parameter would be $\lambda \approx 0.157$,
which is extremely large (only $5 \times 10^{-3}$ of all halos have
spin parameters at least that large). However,
this argument may be flawed because the figure rotation
is a pattern speed, not the speed of the individual particles
which constitute the halo, and so it is in principle independent
of the angular momentum; in some cases the figure may even rotate
retrograde to the particle orbits \citep{freeman66}.
\citet{schwarzschild82} discusses in detail the orbits inside
elliptical galaxies with figure rotation.
He finds that models can be constructed from box and $X$-tube orbits,
which have no net streaming of particles with respect to the figure
(though they have prograde streaming at small radius and retrograde streaming
at large radius), and so result in figures and particles with the
same net rotation.
He also constructs models that include prograde-streaming $Z$-tube
orbits, which result in a figure that rotates slower than the
particles.
Stable retrograde $Z$-tube orbits also
exist, but \citet{schwarzschild82} did not attempt to include them
in his models, so it may also be possible for the figure to rotate
faster than the particles.
While these results demonstrate the independence of the figure
and particle rotation,
it is not clear if they can be translated directly
to dark matter halos.
Dark matter halos may have different formation mechanisms and may be subject to different
tidal forces than elliptical galaxies,
and the different density profile may also have a large effect
on the viable orbital families \citep{gb85}.

If there are observational consequences to dark halo figure rotation,
such as those found by \citet{bf02}, they can be used as a direct
method to distinguish between dark matter and models such as
MOdified Newtonian Dynamics (MOND)
that propose to change the strength of the force of gravity
\citep{milgrom83,mond-review}.
Many of the traditional methods of deducing
dark matter cannot easily distinguish between the presence of a
roughly spherical dark matter halo and a modified force or inertia law.
However, a major difference between dark matter and MOND is that
dark matter is dynamical, and so tests that detect the presence of
dark matter in motion are an effective tool to discriminate between
these possibilities. Among the tests that can make this distinction
are the ellipticities of dark matter halos as measured
using X-ray isophotes, the Sunyaev-Zeldovich effect, and weak lensing
\citep{buote-etal02,ls03,ls04,hyg04};
the presence of bars with parameters
consistent with being stimulated by their angular momentum exchange
with the halo \citep{athanassoula02,vk03}; and spatial offset
between the baryons and the mass in infalling substructure measured
using weak lensing \citep*{cgm04}.
Rotation of the halo figure requires that dark matter is dynamic,
and therefore observable structure triggered by figure rotation potentially
provides another test of the dark matter paradigm.

In this paper, we use cosmological simulations to determine
how the figures of \lcdm\ halos rotate.
The organization of the paper is as follows.
Section~\ref{simulation section} presents the cosmological simulations.
Section~\ref{calculation section} describes the method used to calculate
the figure rotations, which are presented in Section~\ref{results section}.
Finally we discuss our conclusions in Section~\ref{conclusions section}.

\section{The Simulations}\label{simulation section}

The halos are drawn from a large high resolution cosmological N-body
simulation performed using the GADGET2 code
\citep{gadget}.
We adopt a ``concordance'' cosmology
\citep[e.g.][]{wmap} with $\Omega_0=0.3$,
$\Omega_{\Lambda}=0.7$, $\Omega_{\mathrm{bar}}=0.045$,
$h=0.7$, and $\sigma_8 = 0.9$.
The only effect of $\Omega_{\mathrm{bar}}$ is on the 
initial power spectrum, since no baryonic physics is included
in the simulation.
The simulation contains
$512^3 = 134,217,728$ particles in a 
periodic volume $50~h^{-1}$~Mpc comoving
on a side.
All results are scaled into $h$-independent units when possible.
The full list of parameters
is given in Table~\ref{cosmological simulation parameters}.

\begin{deluxetable}{lc}
\tablecaption{Parameters of the cosmological simulation.\label{cosmological simulation parameters}}
\tablewidth{0pt}
\tablehead{\colhead{Parameter} & \colhead{Value}}
\startdata
$N$ & $512^3$\\
Box size ($h^{-1}~\mathrm{Mpc}$ comoving) & $50$\\
Particle mass ($10^7~\hmsun$) & 7.757\\
Force softening length ($h^{-1}~\mathrm{kpc}$) & 5\\
Hubble parameter $h$ ($H_0 = 100~h~\mathrm{km~s^{-1}~Mpc^{-1}}$) & 0.7\\
$\Omega_0$ & 0.3\\
$\Omega_\Lambda$ & 0.7\\
$\sigma_8$ & 0.9\\
$\Omega_{\mathrm{bar}}$ & 0.045\\
\enddata
\end{deluxetable}

A friends-of-friends algorithm is used to identify halos
\citep{defw85}.
We use the standard linking length of
\begin{equation}\label{eq linking length}
b = 0.2 \bar{n}^{-1/3},
\end{equation}
where $\bar{n} = N/V$ is the global number density.

Measuring the figure rotation requires comparing the same halo at different
times during the simulation. We analyze
snapshots of the simulation at lookback times of
approximately 1000, 500, 300, and 100~\hmyr\ with respect to
the $z=0$ snapshot. The scale factor $a$
of each snapshot, along with its corresponding redshift and
lookback time, is listed in Table~\ref{snapshot redshift list}.

\begin{deluxetable}{lllc}
\tablecaption{Snapshots used to calculate
figure rotations.\label{snapshot redshift list}}
\tablewidth{0pt}
\tablehead{\colhead{Snapshot Name} & \colhead{Scale Factor} &
\colhead{Redshift} & \colhead{Lookback Time}\\
 & \colhead{($a$)} & \colhead{($z$)} & \colhead{(\hmyr)}}
\startdata
b090 & 0.89 & 0.1236 & 1108\\
b096 & 0.95 & 0.0526 & 496\\
b098 & 0.97 & 0.0309 & 296\\
b100 & 0.99 & 0.0101 & 98\\
b102 & 1.0 & 0.0 & 0\\
\enddata
\end{deluxetable}

\section{Methodology}\label{calculation section}

\subsection{Introduction}\label{methodology intro section}

The basic method is to identify individual halos in the final
$z=0$ snapshot of the simulation, to find their respective progenitors in slightly
earlier snapshots, and to measure the rotation of the major axes
through their common plane as a function of time.

Precisely determining the direction of the axes is crucial
and difficult. When merely calculating
axial ratios or internal alignment, errors on the order of a few degrees
are tolerable. However, if a pattern speed of 1~\kmskpc, as observed in the halo of \citetalias{bureau-etal99}, is typical, then a typical
halo will only rotate by 4\degr\ in between the penultimate and final snapshots of
the simulation. Therefore, 
the axes of a halo must be determined more precisely than this
in order for the figure rotation to be detectable.
In fact, we should strive for even smaller errors
to see if slower-rotating halos exist.
It would have been difficult for \citetalias{pfitzner99}
to detect halos rotating much slower
than the halo presented in \citetalias{bureau-etal99};
although the error varies from halo to halo (for reasons
discussed in section~\ref{axis error section}),
Figure~5.23 of \citetalias{pfitzner99} shows that most of his halos had
orientation errors of between
8\degr\ and 15\degr, corresponding to a minimum resolvable figure rotation
of $\sim 0.6~\kmskpc$ for a $2\sigma$ detection in snapshots spaced
500~Myr apart.

A major difficulty in determining the principal axes so precisely
is substructure. The orientation of a mass distribution is
usually found by calculating the moment of inertia tensor 
$I_{ij} = \sum_k m_k r_{k,i} r_{k,j}$, and then diagonalizing $I_{ij}$
to find the principal axes. However, this procedure weights particles
by $r^2$. Therefore,
substructure near the edge of the halo (or of the subregion of the halo
used to calculate the shape) can exert a large influence on the shape
of nearly spherical halos, especially if a particular subhalo is part of
the calculation in one snapshot but not in another, such as when
it has just fallen in.
This is particularly problematic because subhalos are preferentially
found at large radii \citep{ghigna-etal00,delucia-etal04,gkg04,gao-etal04}.
Moving substructures can also induce a false measurement of figure
rotation due to their motion within the main halo at approximately
the circular velocity.

To mitigate this, we firstly use particles in a spherical region
of radius 0.6~\rvir\ 
surrounding the center of the halo, rather than picking the particles
from a density dependent ellipsoid as in \citet{warren-etal92}
or \citet{js02}.
We find that those
methods allow substructure at one particular radius to influence
the overall shape of the ellipsoid from which particles are chosen
for the remainder of the calculation, and therefore bias the results
even when other measures are adopted to minimize their effect.
The choice of a spherical region biases the derived axis
ratios toward spherical values, but does not affect the
orientation. Secondly,
the particles are weighted by $1/r^2$ so that each mass unit contributes
equally regardless of radius \citep{gerhard83}.
Both \citetalias{dubinski92}
and \citetalias{pfitzner99} take similar approaches, but using
radii based on ellipsoidal shells. Therefore, we base our analysis on 
the principal axes of the reduced
inertia tensor 
\begin{equation}\label{eq inertia tensor}
\tilde{I}_{ij} = \sum_k \frac{ r_{k,i} r_{k,j} }{r_k^2}.
\end{equation}
In the majority of halos, the
substructure is a small fraction of the total mass, usually less than
5\%\ of the total mass within 60\%\ of the virial radius
\citep[Figure~8]{delucia-etal04},
so its effect is much reduced.
There are still some halos which
have not yet relaxed from a recent major merger, in which case
the ``substructure'' constitutes a significant fraction of the halo.
To find these cases, the reduced inertia tensor is
separately calculated enclosing
spheres of radius 0.6, 0.4, 0.25, 0.12, and 0.06 times the
virial radius to look for deviations as a function of radius
(see section~\ref{5sigma deviations} for details).
These radii are always with respect to the $z=0$ value of \rvir.

We find that only halos with at least
$4 \times 10^3$ particles, or
masses of at least $\sim 3 \times 10^{11}~\hmsun$
have sufficient resolution for the orientation of the principal axis
to be determined at sufficient precision
 (see Section~\ref{axis error section}).
There are 1432 halos in the $z=0$ snapshot satisfying this criterion,
with masses extending up to $2.8 \times 10^{14}~\hmsun$.

\subsection{Halo matching}\label{halo matching section}

\begin{figure}
\plotone{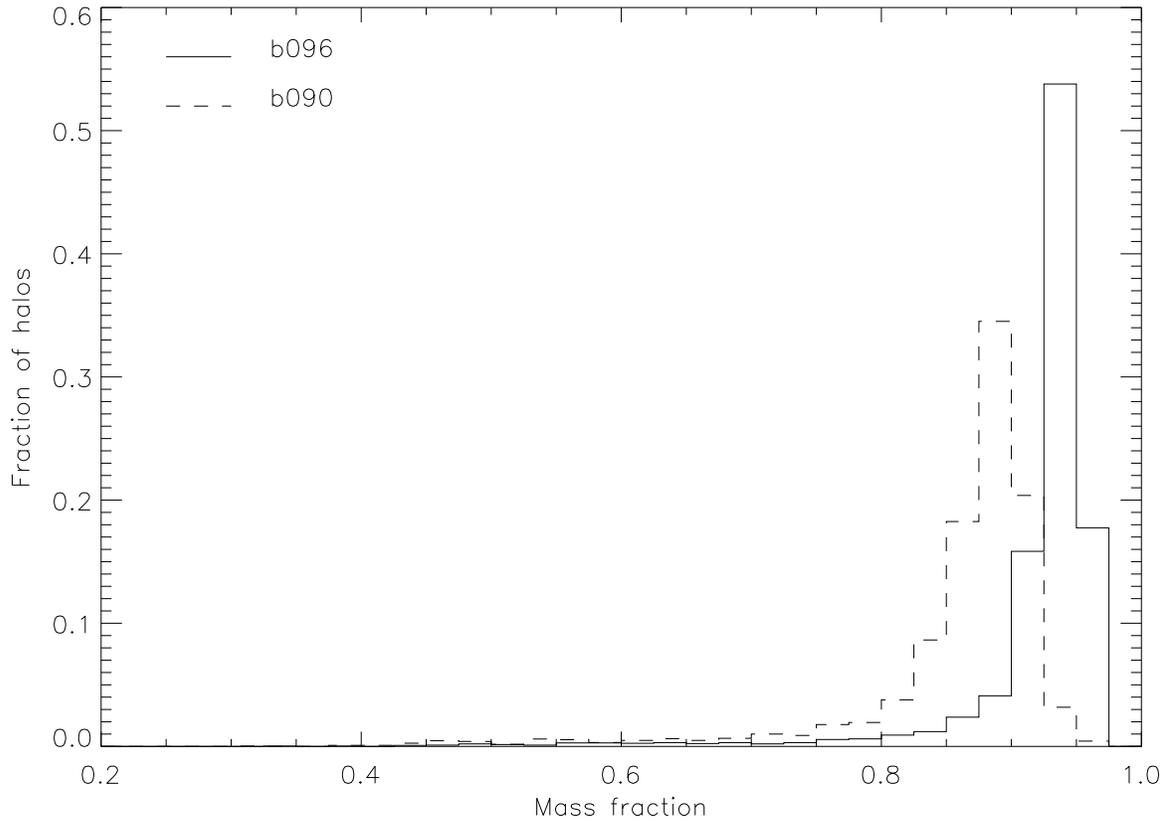}
\caption{\label{progenitor mass fraction plot} 
Histogram of the fraction of the final mass that comes from the
b096 ($z \approx 0.05$, solid line)
and b090 ($z \approx 0.12$, dashed line)
halo which contributes the most mass.%
}
\end{figure}

To match up the halos at $z=0$ with their earlier counterparts,
we use the individual particle numbers provided by GADGET which
are invariant from snapshot to snapshot, and
find which halo each particle belongs to in each snapshot.
The progenitor of each $z=0$ halo in a given $z>0$
snapshot is the halo that contributes $\ge 90\%$ of the final
halo mass. Sometimes no such halo exists;
in these cases, the
halo has only just formed or underwent a major merger and so is not useful
for our purposes.
Figure~\ref{progenitor mass fraction plot} shows a histogram of the fraction
of the final halo mass that comes from the b096 ($z \approx 0.05$)
halo which contributes the most mass.
There are also some cases where two nearby objects are
identified as a single halo in an earlier snapshot but as distinct
objects in the final snapshot. We therefore impose the additional constraint
that the mass contributed to the final halo must also be $\ge 90\%$ of the
progenitor's mass.
In the longer time between the earliest snapshot b090 and the final
snapshot b102, a halo typically accretes a greater fraction of its
mass, and so a more liberal cut of 85\%\ is used for this snapshot
(see the dashed histogram in Figure~\ref{progenitor mass fraction plot}).
492 of the halos that satisfied the mass cut
did not have a progenitor which satisfied these criteria in at least
one of the $z>0$ snapshots and so were eliminated from the analysis,
leaving a sample of 940 matched halos.

\subsection{Error in axis orientation}\label{axis error section}

There are two sources of errors that enter into the determination of the
axes: how well
the principal axes of the particle distribution can be determined,
and whether that particle distribution has a smooth triaxial figure.
Here we estimate the error
assuming that it is not biased by substructure.
The halos for which this assumption does not hold
will become apparent later in the calculation.

For a smooth triaxial ellipsoid represented by $N$ particles,
the error is a function of $N$ and of the
intrinsic shape: as the axis ratio
$b/a$ or $c/b$ approaches unity, the axes become degenerate.
To quantify this, we have performed a bootstrap analysis of the particles
in a sphere of radius 0.6~\rvir\ of each $z=0$ halo \citep{heyl-etal94}.
If the sphere contains $N$ particles then we
resample the structure by
randomly selecting $N$ particles from that set
allowing for duplication and determine the axes from this bootstrap
set. We do this 100 times for each halo. The dispersion
of these estimates around the calculated axis is 
taken formally as the ``$1\sigma$'' angular error, and is labelled
\thboot.

\begin{figure}
\plotone{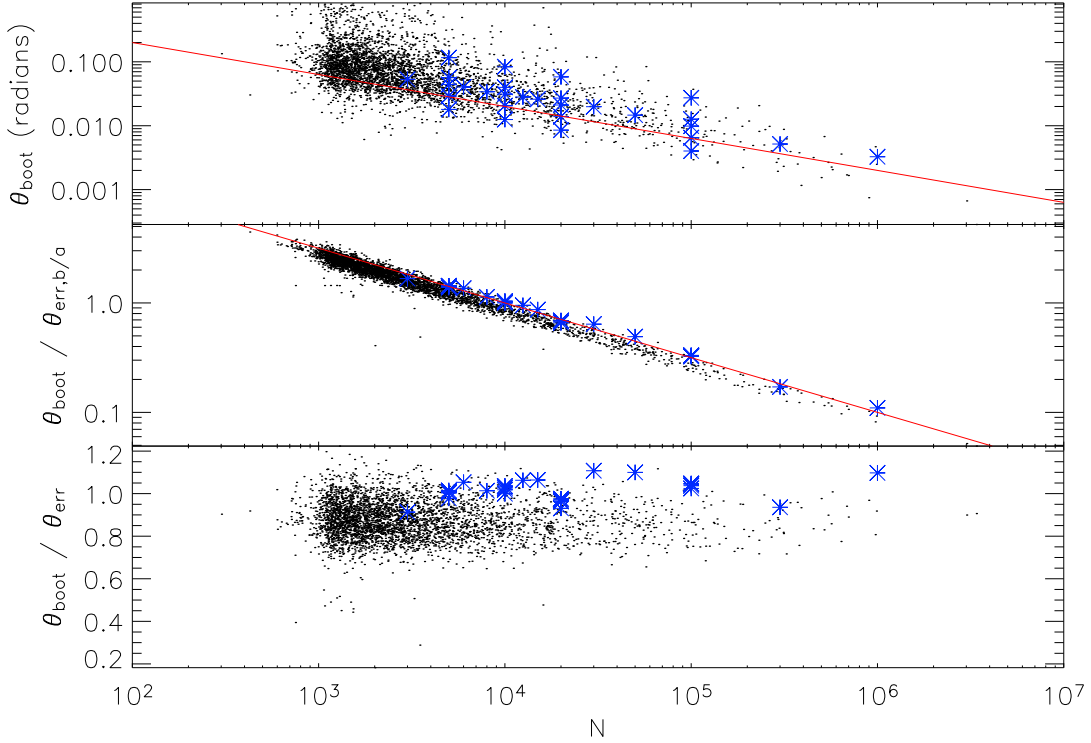}
%\plotone{f2-bw.eps}
\caption{\label{major axis error vs N}
Angular bootstrap error \thboot\ as a function of the number
of particles $N$ within the central 0.6~\rvir\ of each halo.
Points are the cosmological halos, and asterisks are
randomly sampled smooth NFW halos.
\textit{(Top):} Angular error \thboot. The solid line
is the fit \therrN\ from equation~(\ref{theta err N}).
\textit{(Middle):} Ratio between the angular error and the
error expected for the halo given its axis ratio $b/a$, i.e.
$\thboot / \therrba$.
The solid line is \therrN\ from equation~(\ref{theta err N})
renormalized to the typical error of 0.02~radians.
\textit{(Bottom):} Ratio between the angular error and the analytic estimate
\therr\ from equation~(\ref{predicted error}).
}
\end{figure}

\begin{figure}
\plotone{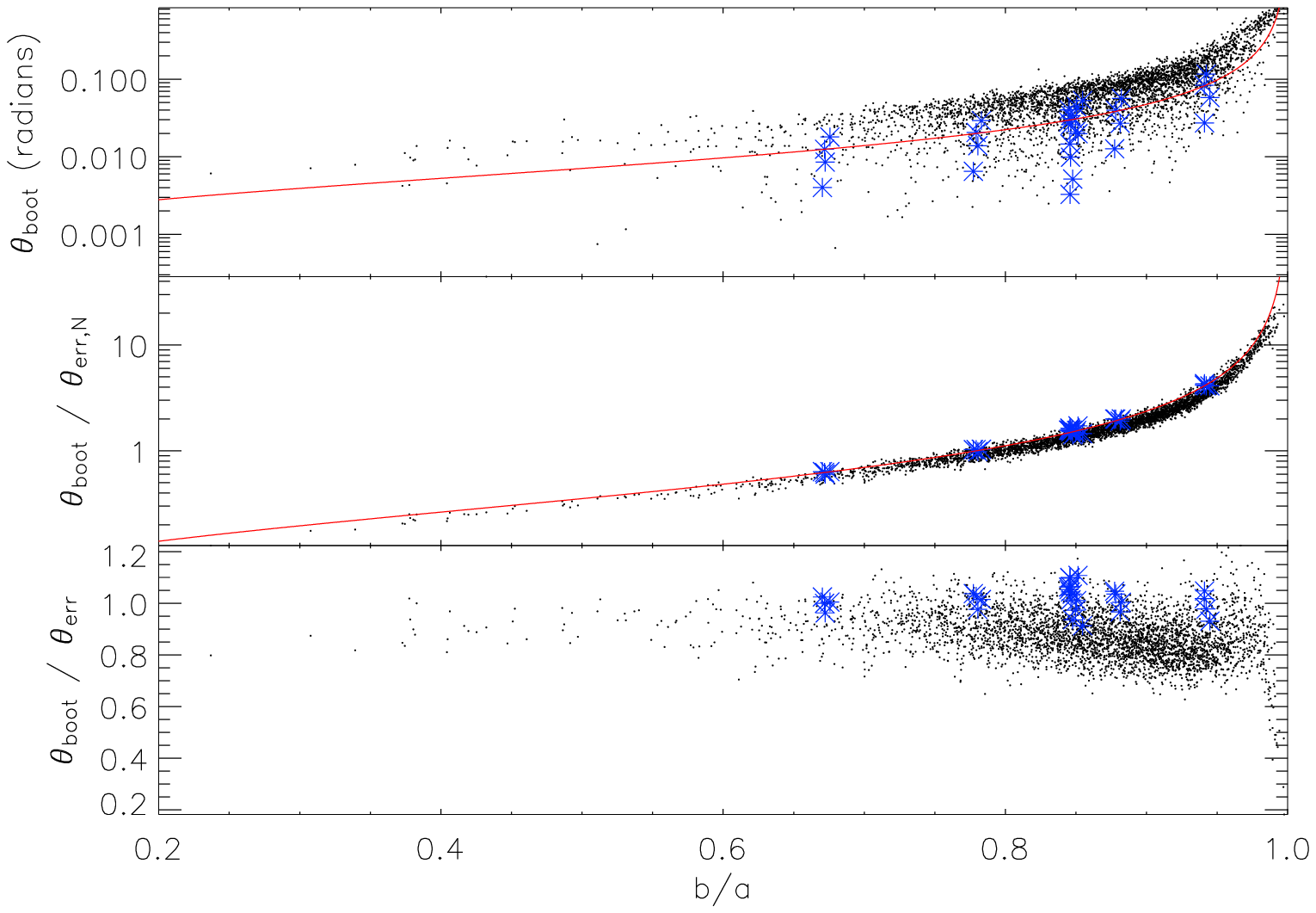}
%\plotone{f3-bw.eps}
\caption{\label{major axis error vs b/a}
Angular bootstrap error \thboot\ as a function of the
axis ratio $b/a$ of each halo.
Points are the cosmological halos, and asterisks are
randomly sampled smooth NFW halos.
\textit{(Top):} Angular error \thboot. The solid line
is the fit \therrba\ from equation~(\ref{theta err b/a}).
\textit{(Middle):} Ratio between the angular error and the
error expected for the halo given the number of particles $N$, i.e.
$\thboot / \therrN$.
The solid line is \therrba\ from equation~(\ref{theta err b/a})
renormalized to the typical error of 0.02~radians.
\textit{(Bottom):} Ratio between the angular error and the analytic estimate
\therr\ from equation~(\ref{predicted error}).
}
\end{figure}

As expected, the two important parameters are
$N$ and the axis ratio. We focus here on the major axis, for which the
important axis ratio is $b/a$.
The top panels of Figures~\ref{major axis error vs N}
and~\ref{major axis error vs b/a} show the dependence of the
bootstrap error on $N$ and $b/a$ respectively
for all halos with $M > 10^{11}~\hmsun$.
The solid lines are empirical fits:
\begin{equation}\label{theta err N}
\therrN = \frac{2}{\sqrt{N}},
\end{equation}
and
\begin{equation}\label{theta err b/a}
\therrba = 0.005 \frac{\sqrt{b/a}}{1 - b/a}.
\end{equation}
The form of equation~(\ref{theta err N}) is not surprising; if a smooth
halo was randomly sampled, we would expect the errors to be
Poissonian with an $N^{-1/2}$ dependence. However, the
cosmological halos are not randomly sampled. Individual particles ``know''
where the other particles are, because they have acquired their positions
by reacting in the potential defined by those other particles.
Therefore, the errors may be less than expected from a randomly
sampled halo. To test this, we construct a series of smooth prolate NFW
halos \citep*{nfw96} with $b/a$ axis ratios ranging from 0.5~to 0.9,
randomly sampled with between $3\times 10^3$ and $3\times 10^5$ particles,
and perform the bootstrap analysis identically for each of these halos
as for the
cosmological halos.
Because the method for calculating axis ratios outlined in
Section~\ref{methodology intro section} biases axis ratios toward
spherical, the recovered $b/a$ of these randomly sampled halos
is larger than the input value, and ranges from 0.65~to 0.95.
The errors for these randomly sampled smooth halos are shown as asterisks in
Figures~\ref{major axis error vs N} and~\ref{major axis error vs b/a}.

The top panel of Figure~\ref{major axis error vs N} shows a rise in the
dispersion of the error for $N\lesssim 4000$, with many halos having
errors greater than the 0.1~radians necessary to detect the figure
rotation of the halo presented in \citetalias{bureau-etal99}.
Therefore, we only use halos with $N>4000$.

The bootstrap error appears to be completely determined by $N$ and
$b/a$. The residuals of \thboot\ with respect to \therrN\ are due
to \therrba\ and vice versa.
This is shown in the middle panels of Figures~\ref{major axis error vs N}
and \ref{major axis error vs b/a}. In
the middle panel of Figure~\ref{major axis error vs N}
we have divided out the dependence of \thboot\ on the axis ratio,
making apparent an extremely tight relation between the residual
and $N$, while in
the middle panel of Figure~\ref{major axis error vs b/a} we have divided
out the dependence of \thboot\ on $N$, showing the equally tight
relation between the residual and $b/a$. It is apparent
from comparing the points and asterisks that
the errors in the cosmological halos are slightly smaller than for randomly
sampled smooth halos.

Combining equations~(\ref{theta err N}) and~(\ref{theta err b/a}), and
noting that the typical error is $\thboot \approx 0.02$~radians, we
find the bootstrap error is well fit by
\begin{equation}\label{predicted error}
\therr = \frac{1}{2 \sqrt{N}} \frac{\sqrt{b/a}}{1 - b/a}.
\end{equation}
The bottom panels of Figures~\ref{major axis error vs N}
and~\ref{major axis error vs b/a} show the residual ratio between
the bootstrap error \thboot\ and the analytic estimate
\therr. The vast majority of points lie between 0.8~and 1.0,
indicating that \therr\ overestimates the error by $\sim 10\%$.
Equation~(\ref{predicted error}) breaks down 
as $b/a$ approaches unity; these halos are nearly oblate and so
do not have well-defined major axes.
It also becomes inaccurate at very low $b/a$ due to low-mass
poorly-resolved halos.
Even in these cases, the error estimate is conservative,
but to be safe we have eliminated axes with $b/a < 0.35$ or $b/a > 0.95$
from the subsequent analysis, regardless of the nominal error.
The randomly-sampled smooth halos follow
equation~(\ref{predicted error}) extremely well, so the
non-Poissonianity of the sampling in simulated halos reduces the errors by 10\%.

Calculating the bootstraps is computationally expensive, so
equation~(\ref{predicted error}) is used for the error
in all further computation. Because this estimate is expected to be
correct for smooth ellipsoids, cases where the error is
anomalous are indications of residual substructure.

\subsection{Figure rotation}\label{figure rotation methodology section}

\begin{figure}
\plotone{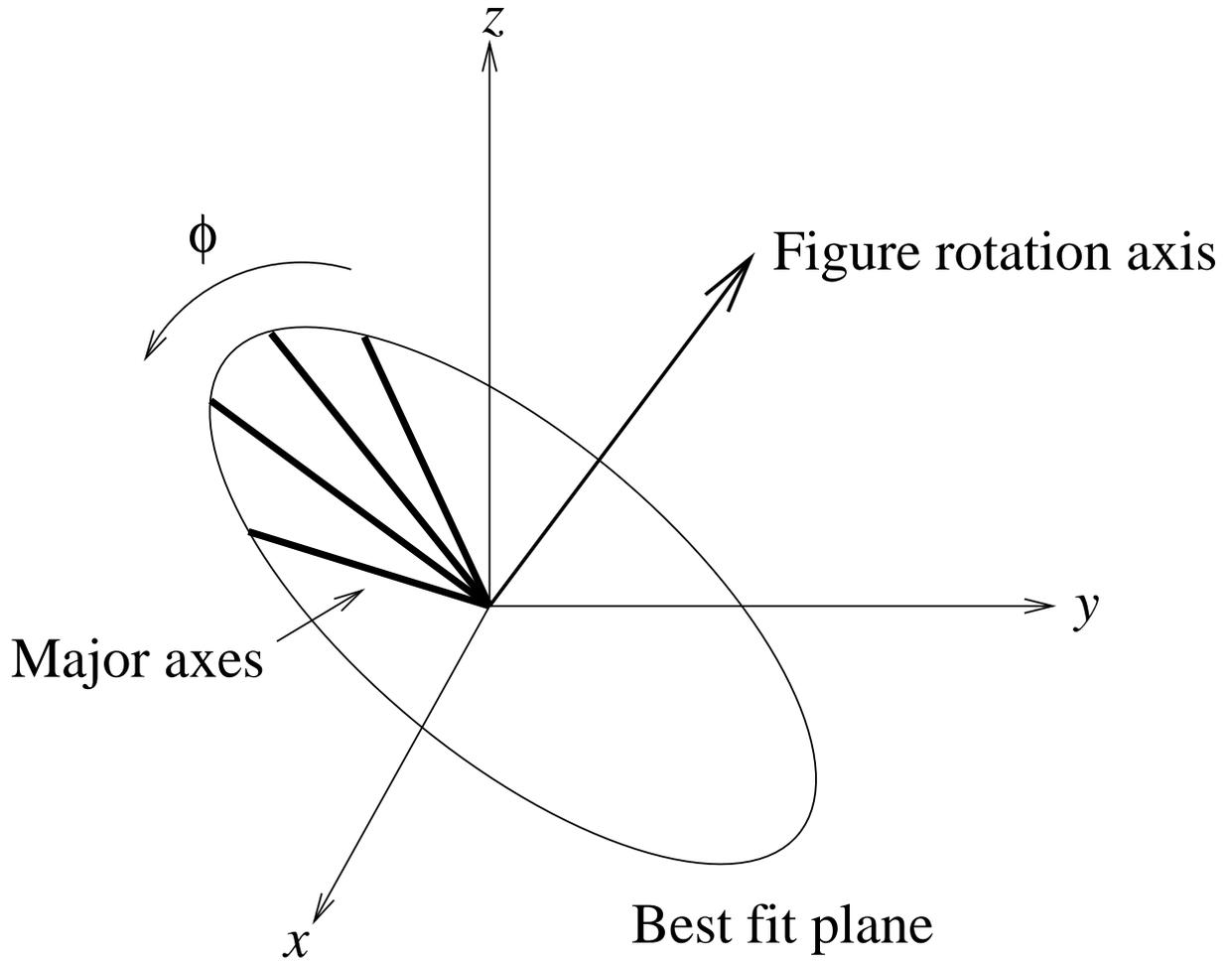}
\caption{\label{plane diagram}%
Diagram that demonstrates how we fit a plane to the major axis
measurements at all snapshots (thick lines) and then find the
increase of phase $\phi$ as a function of time.
The figure rotation axis is perpendicular to the best fit plane,
and defined such that $\phi$ increases around it counter-clockwise
with time.}
\end{figure}

Ideally one would fit the figure rotation by comparing the orientation
of the axis at each snapshot
to that of a unit vector rotating uniformly along a great circle, and minimize
the $\chi^2$ to find the best fit uniform great circle trajectory.
However, this requires minimizing a non-linear function in
a four-dimensional parameter space, a non-trivial task.
We adopt the simpler and numerically more robust method
of solving for the plane
$z = ax + by$ that fits the major axis measurements of the halo
best at all timesteps,
assuming the error is negligible. The change of the phase of the axes
in this plane as a function of time are then fit by
linear regression. A schematic diagram of this process is
shown in Figure~\ref{plane diagram}.

The degree to which the axes are consistent with lying in a plane is
checked by calculating the out-of-plane $\chi^2$:
\begin{equation}\label{out of plane chi2}
\chi^2_{\mathrm{oop}} \equiv 
  \frac{1}{\nu} \sum_i \frac{ \Delta\theta_i^2 } {{\therr}_i^2},
\end{equation}
where $\nu$ is the number of degrees of freedom and
$\Delta\theta_i$ is the minimum angular distance between the
major axis at timestep $i$ and the great circle defined by the best
fit plane.

Because the axes have reflection symmetry, it is impossible to
measure a change in phase of more than $\pi/2$.
The phases are adjusted by units of $\pi$ such that the difference
in phase between adjacent snapshots is always less than $\pi/2$.
If the figure were truly rotating by 90\degr\ or more in between the snapshots,
it would be impossible to accurately measure this rotation
since the angular frequency
would be larger than the Nyquist frequency of our sampling rate.
Any faster pattern speeds would be aliased to lower
angular velocities, with an aliased angular velocity of
$\Omega_{\mathrm{Nyq}} - (\Omega_p - \Omega_{\mathrm{Nyq}})$,
where $\Omega_p$ is the intrinsic angular velocity of the pattern
and $\Omega_{\mathrm{Nyq}}$ is the Nyquist frequency.
For snapshots spaced $500~h^{-1}~\mathrm{Myr}$ apart, the
maximum time between the snapshots we analyze,
the maximum detectable angular velocity is 3.8~\hkmskpc.
We do not expect the figure to change so dramatically as we have excluded
major mergers. However, this can be checked \textit{post facto} by checking
whether the distribution of measured angular velocities extends up to
the Nyquist frequency;
if so, then there are likely even more rapidly rotating figures whose
angular frequency
is aliased into the detectable range, fooling us into thinking they are
rotating slower.
If the measured distribution does
not extend to the Nyquist frequency, then it is unlikely that
there are any figures rotating too rapidly to be detected
(see Section~\ref{results section}).

The best fit linear relation for the phase as a function of time is found
by linear regression. Because the component of an isotropic angular error
projected onto a plane is half of the isotropic error, we divide
the error of equation~(\ref{predicted error}) by two before we perform the 
regression.
The slope of the linear fit gives the
pattern speed $\Omega_p$ of the figure rotation. The error is the
$1\sigma$ limit on the slope.

\begin{figure}
\plotone{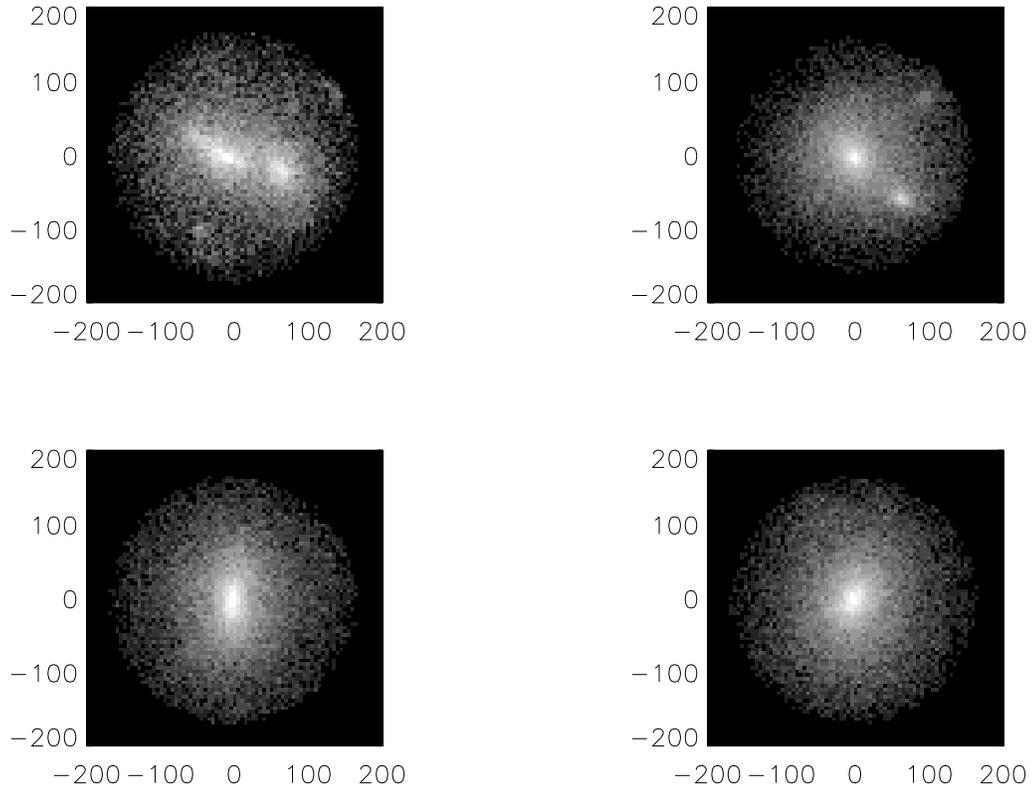}
\caption{\label{subhalo fraction}
Log-weighted projected density of 4 halos with a range of subhalo fractions
$f_s$. The subhalo fractions are 0.166 (top-left), 0.065 (top-right),
0.045 (bottom-left), and 0.016 (bottom-right).
Axes are in units of $h^{-1}~\mathrm{kpc}$ from the halo center.
All halos have masses in the
range $2$ -- $3 \times 10^{12}~\hmsun$.
}
\end{figure}

\label{5sigma deviations}%
Once we have calculated the pattern speed for each halo, we can
eliminate the cases where substructure has severely impacted the
results. In these cases, the signal is dominated by a large subhalo
at a particular radius, so the derived pattern speed will be significantly
different when the sphere is large enough to include the subhalo
from when the subhalo is outside the sphere. We have calculated the
pattern speed using enclosing spheres of radius 0.6, 0.4, 0.25,
0.12, and 0.06 of the virial radius.
The fraction of mass in subhalos can be estimated via the
change in the pattern speed $\Omega_p$ at adjacent radii.
Because the reduced inertia tensor is mass-weighted, the
figure rotation of a sphere
with a smooth component rotating at $\Omega_{p,\mathrm{smooth}}$
plus a subhalo containing a fraction $f_s$ of the total mass
moving at the circular velocity $v_c$ at radius
$R~\rvir$ is approximately
\begin{equation}\label{omegap for a subhalo}
\Omega_p \approx (1-f_s) \Omega_{p,\mathrm{smooth}} + f_s \frac{v_c}{R~\rvir},
\end{equation}
where the difference due to the presence of the subhalo is
\begin{equation}\label{delta omegap = fs vc / r}
\Delta \Omega_p = f_s \frac{v_c}{R~\rvir} = f_s \sqrt{
  \frac{ G M(<R~\rvir) }{R^3~\rvir^3}}.
\end{equation}
If the density profile is roughly isothermal, the enclosed mass is
\begin{equation}\label{eq enclosed mass}
M(<R~\rvir) = \frac{4}{3} \pi \Delta_c \rhocrit R~\rvir^3.
\end{equation}
Solving equations~(\ref{delta omegap = fs vc / r}) and~(\ref{eq enclosed mass})
gives an expression for
the fraction of the mass in substructure given a jump of $\Delta \Omega_p$
in the measured pattern speed when crossing radius $R~\rvir$:
\begin{equation}\label{fs solution eq}
f_s = \frac{ \Delta \Omega_p R } { \sqrt{\frac{4}{3} \pi G \Delta_c \rhocrit}}.
\end{equation}
The term in the square root is equal to $0.72~\hkmskpc$.
For each halo, we compute the value of $f_s$ due to the jump
$\Delta \Omega_p$ between each set of adjacent radii, i.e.~for
$R=0.4, 0.25, 0.12,$ and 0.06, and adopt the largest value of $f_s$
as the substructure fraction of the halo.
Figure~\ref{subhalo fraction} shows log-weighted projected densities of 4
halos in the mass range $2$ -- $3 \times 10^{12}~\hmsun$
with a variety of values of $f_s$, ranging from 0.166 at the top-left
to 0.016 at the bottom right. After examining a number of halos spanning
a range of $f_s$, we adopt a cutoff of $f_s < 0.05$ for undisturbed
halos.
This eliminates 289 of the 940 halos, leaving 651 undisturbed halos.

A further 158 halos were eliminated because the angular error
approached $\pi/2$ in at least one of the snapshots.
This includes the halos with $b/a<0.35$ or $b/a>0.95$ discussed in
Section~\ref{axis error section}.
We also eliminate cases where
the reduced $\chi^2$ from the linear fit of phase versus time
indicates that the intrinsic error of the direction
determination is much lower than suggested by
equation~(\ref{predicted error}), indicating that the model of the halo
as a smooth ellipsoid is violated (10 halos
with $\chi^2_{\nu} < 0.1$),
and those cases where the phase does not evolve linearly with
time (134 halos with $\chi^2_{\nu} > 10$).
Finally, we eliminate halos where the axes do not lie on a common
plane, i.e. the 32 halos where $\chi^2_{\mathrm{oop}} > 10$.
Therefore, the final sample consists of 317~halos.

\begin{figure}
\plotone{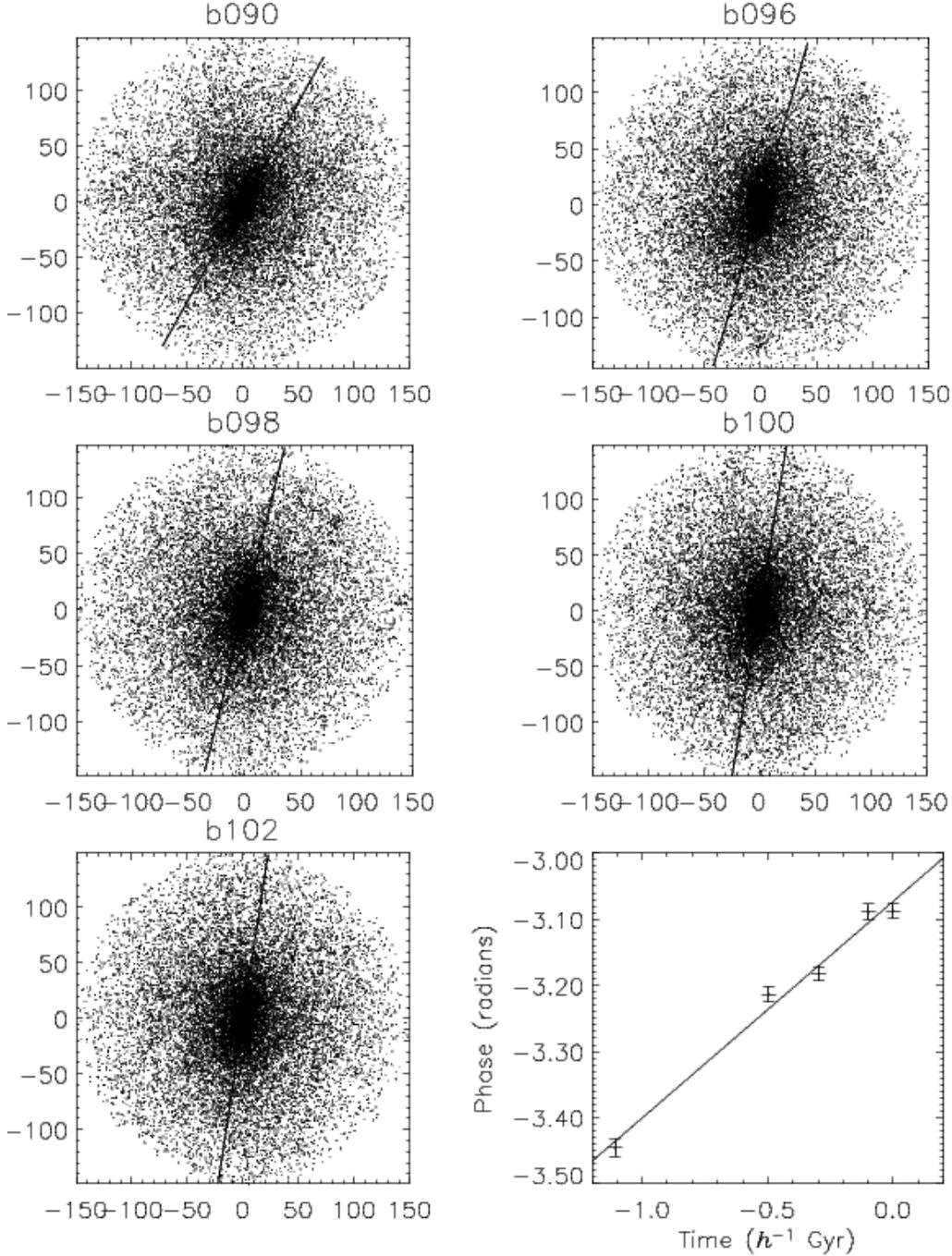}
\caption{\textit{(Upper-left five panels):}
Projection onto the best fit plane
of the inner 0.6~\rvir\ of a sample halo at the five snapshots
we analyze.
Axes are in units of $h^{-1}~\mathrm{kpc}$ from the halo center.
From left to right, top to bottom, the snapshots are at
1108, 496, 296, 98, and 0~\hmyr\ before $z=0$.
The solid line is the major axis, which rotates counterclockwise
by 20\degr\ from beginning to end.
\textit{(Bottom-right):} Phase of the major axis in the rotational
plane of the sample halo.
The zero point is arbitrary, but identical in all snapshots.
The solid line is the linear
fit, with a slope of $0.33~h~\mathrm{radians~Gyr^{-1}}$.
\label{sample halo}}
\end{figure}

A sample halo is shown in
the first five panels of Figure~\ref{sample halo}. It was chosen
randomly from the halos with relatively low errors and typical
pattern speeds. It has a mass of $1.9\times 10^{12}~\hmsun$,
and a pattern that rotates at $0.32\pm 0.01~\hkmskpc$.
It has a spin parameter $\lambda = 0.047$, and axis ratios of
$b/a=0.86$ and $c/a=0.77$ at $z=0$. The derived substructure fraction
is $f_s=0.045$, and the out-of-plane $\chi^2_{\mathrm{oop}} = 8.5$.
The solid line shows the measured major axis in each
snapshot, which rotates counterclockwise in this projection.
The phase of its figure rotation as a function of time is shown in
the bottom-right panel of Figure~\ref{sample halo}.
The zero point is arbitrary,
but is consistent from snapshot to snapshot.
The linear fit is also shown, which has a reduced $\chi^2$ of
2.9.

\section{Results}\label{results section}

\begin{figure}
\plotone{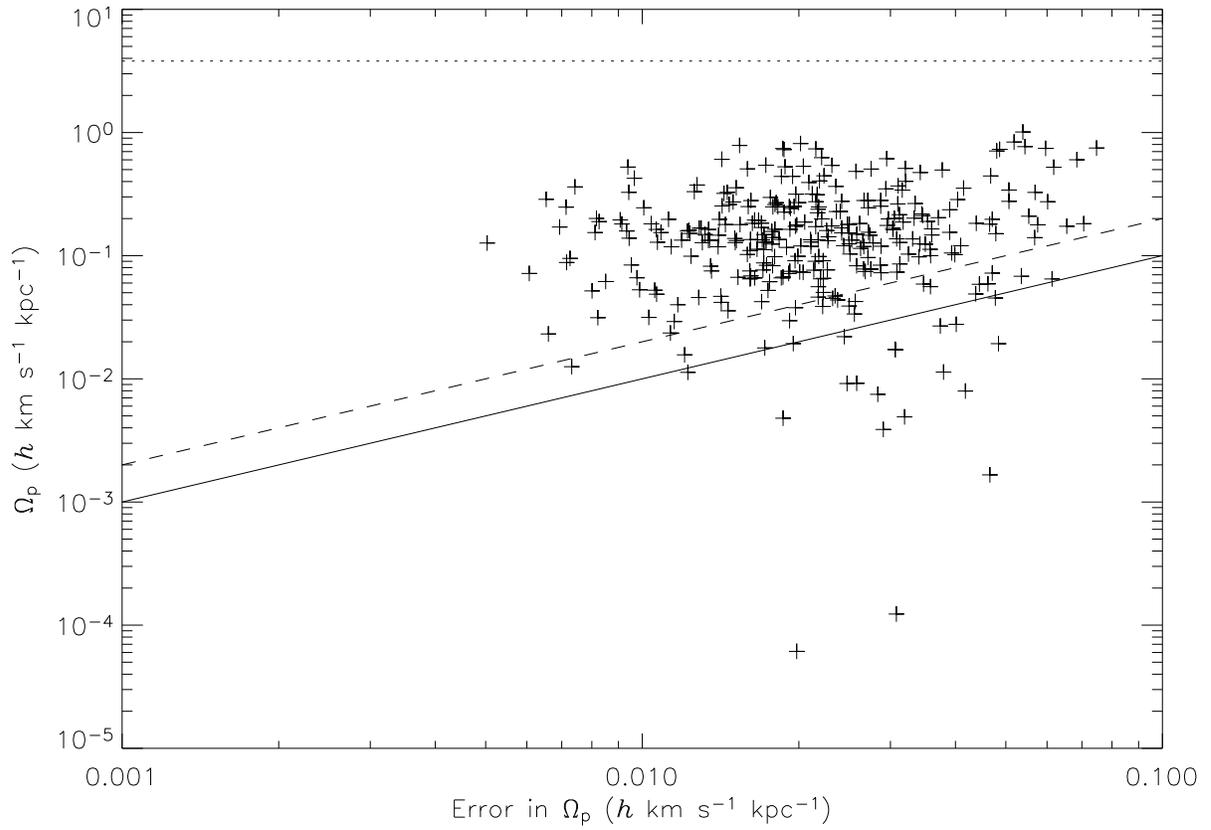}
\caption{Measured pattern speed of the rotation of the major axis.
The $x$-axis is the error in the pattern speed.
The solid line shows where the measure pattern speed is equal
to the estimated error,
while the dashed line is the $2\sigma$ limit. The horizontal
dotted line shows the Nyquist limit.%
\label{omp vs err}}
\end{figure}

\begin{figure}
\plotone{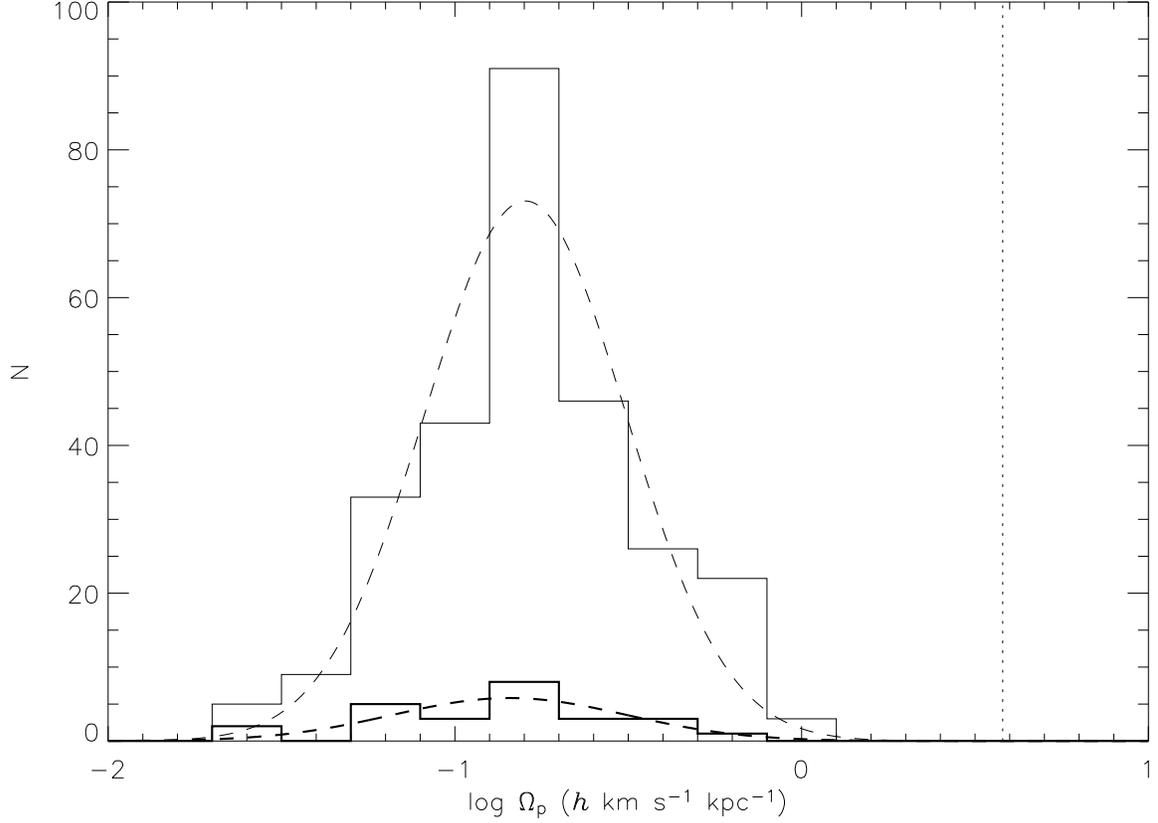}
\caption{Histogram of the pattern speeds of the figure rotation,
expressed in $\log \Omega_p$.
The thin histogram contains all halos that have $2\sigma$
detections of figure rotation, i.e. those above the dashed line
of Figure~\ref{omp vs err}, and is incomplete at $\Omega_p < 0.126~\hkmskpc$
or equivalently $\log \Omega_p < -0.9$. The thick histogram contains
only those halos with errors less than 0.01~\hkmskpc, and is
incomplete at $\Omega_p < 0.015~\hkmskpc$ or $\log \Omega_p < -1.8$.
The dashed curves are Gaussian fits to the histograms.
The vertical dotted line shows the Nyquist limit.%
\label{omp hist}}
\end{figure}

\begin{figure}
\plotone{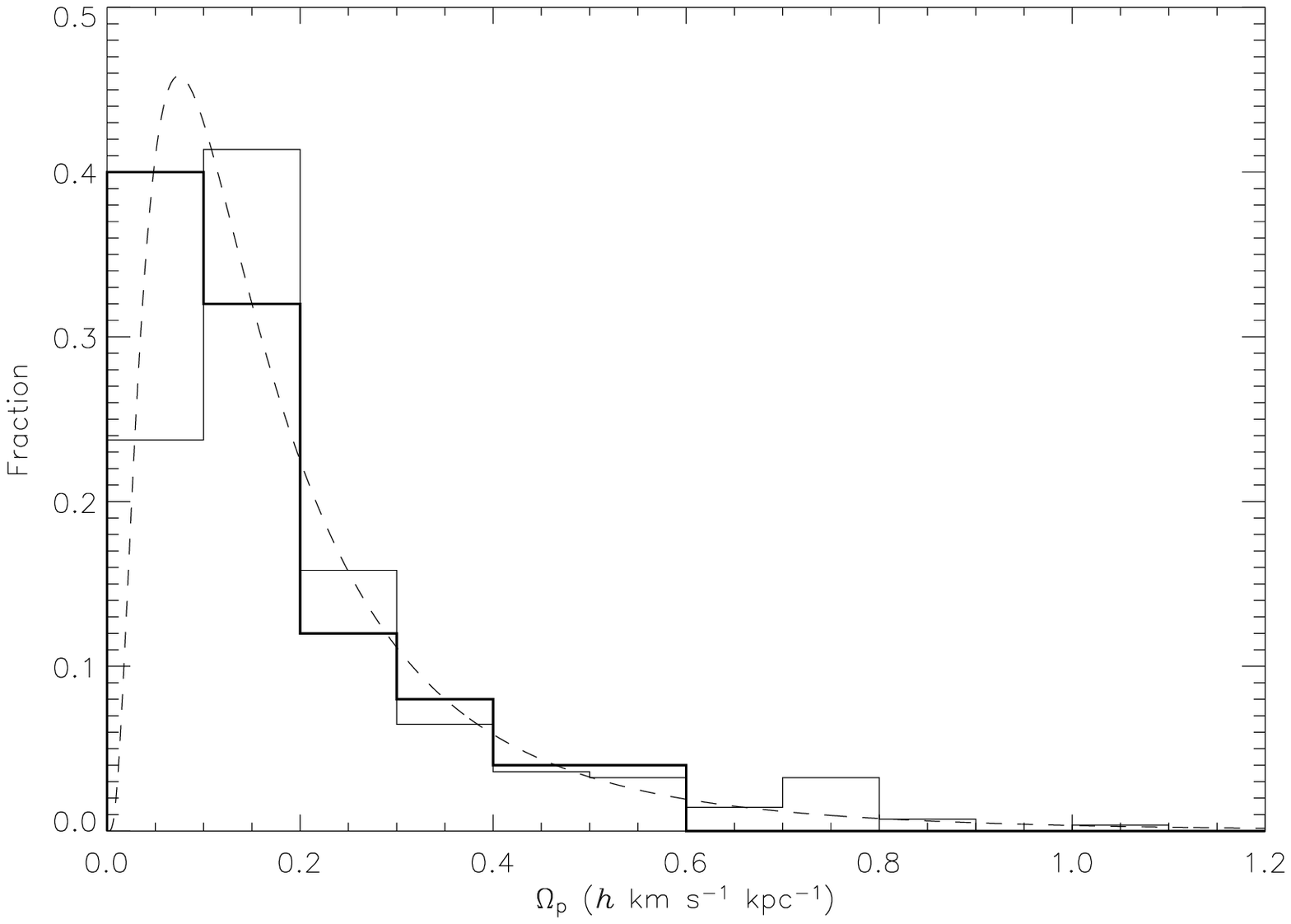}
\caption{Fractional histogram of pattern speeds of the figure rotation,
expressed linearly in $\Omega_p$.
The thin histogram contains all halos that have $2\sigma$
detections of figure rotation (incomplete at $\Omega_p < 0.126~\hkmskpc$),
while the thick histogram contains halos with errors less than
0.01~\hkmskpc and is complete down to $\Omega_p=0.015~\hkmskpc$.
The dashed curve is the log normal fit given by equation~(\ref{log normal}).%
\label{linear fractional histogram}}
\end{figure}

Figure~\ref{omp vs err} shows the measured figure rotation speeds
for all of the halos in the sample, as a function of their error.
Halos with
measured pattern speeds less than twice as large as the estimated error
(the dashed line)
are taken as non-detections. 278 of the 317 halos have detected
figure rotation. A histogram of the pattern speeds is presented in
Figure~\ref{omp hist}, expressed in $\log \Omega_p$.
The thin histogram contains all halos with
$2\sigma$ detections, while the thick histogram contains those with
the smallest errors, less than 0.01~\hkmskpc.
The largest upper limit due to a non-detection
in the main sample is $\Omega_p < 0.126$
($\log \Omega_p < -0.9$), so
the thin histogram is incomplete below this level, while the
low error sample contains only one upper limit, $\Omega_p < 0.015$
($\log \Omega_p < -1.8$), so the thick histogram is complete
down to this level.
The dashed curves are Gaussian fits to
the histograms. The fit to the thin histogram, which has the largest sample
size but is incomplete at low $\Omega_p$, peaks at 
$\log \Omega_p = -0.80$ and has
a standard deviation of 0.29, while the thick curve, which contains fewer halos
but is less biased toward large values of $\Omega_p$,
peaks at $\log \Omega_p = -0.84$ and has a standard deviation of 0.34.
We give more weight to the thick histogram, whose points all have very small
errors, and propose that the true distribution peaks at
$\log \Omega_p = -0.83$ with a
standard deviation of 0.36.
Expressed as a log normal distribution, the probability is
\begin{equation}\label{log normal}
P(\Omega_p) = \frac{1}{\Omega_p \sigma \sqrt{2\pi}}
  \exp\left(-\frac{\ln^2(\Omega_p/\Omega_{p_0})} {2\sigma^2} \right),
\end{equation}
where $\Omega_{p_0} = 10^{-0.83} = 0.148~\hkmskpc$ and the natural width
$\sigma = 0.36 \ln 10 = 0.83$. This fit is shown in
Figure~\ref{linear fractional histogram},
compared to the fractional distribution of
halos in the full (thin) and low error (thick) samples, and encompasses
both the large number of halos with low $\Omega_p$ seen when the errors
are sufficiently small, and the tail at high $\Omega_p$ seen when the sample
size is sufficiently large.

For comparison, the halo in \citetalias{bureau-etal99} has a pattern
speed of 2~\hkmskpc. This lies slightly above the top end of our
distribution; the maximum pattern speed in our sample is 1.01~\hkmskpc.
Based on the log normal fit of equation~(\ref{log normal}),
we estimate the fraction
of halos with $\Omega_p \ge 2~\hkmskpc$ to be $\sim 10^{-3}$. Therefore, this
halo is unusual, but it is not unreasonable to find a halo with
such a pattern speed in a large simulation.
Given the size of the errors in \citetalias{pfitzner99}, and that he
found very few halos with figure rotation, it should not be surprising
that \citetalias{pfitzner99} could only detect pattern speeds at the upper end
of the overall distribution.
The different adopted cosmologies may also influence the results
(note, however, that this comparison is performed in $h$-independent units).
Our results are also mostly consistent with
\citetalias{dubinski92}, who finds pattern speeds of between 0.1~and
1.6~\kmskpc\ in a sample of 14~halos. We have trouble reproducing the most
rapidly rotating halos in \citetalias{dubinski92},
but this may be a product of the heuristic
initial conditions in \citetalias{dubinski92}  compared to the cosmological
initial conditions we use.

In order to account for the spiral structure in \ntnif,
a triaxial figure would need to rotate at $7\pm 1~\kmskpc$ \citepalias{mb03}.
This is almost an order of magnitude faster than the fastest of the halos
in our sample, and the
log normal fit from equation~(\ref{log normal})
suggests that the fraction of halos
with $\Omega_p \ge 6~\kmskpc$ is $5 \times 10^{-7}$.
Therefore, the figure rotation of undisturbed
\lcdm\ halos can not explain the spiral structure of \ntnif.
SPH simulations of gas disks inside triaxial halos with
pattern speeds of 0.77~\kmskpc, comparable to the fastest
pattern speeds in our sample,
show very weak if any enhancement
of spiral structure compared to a static halo \citep[][Figure~2f]{bf02}.
Therefore, it is unlikely that triaxial figure rotation can be
detected in extended gas disks.

\begin{figure}
\plotone{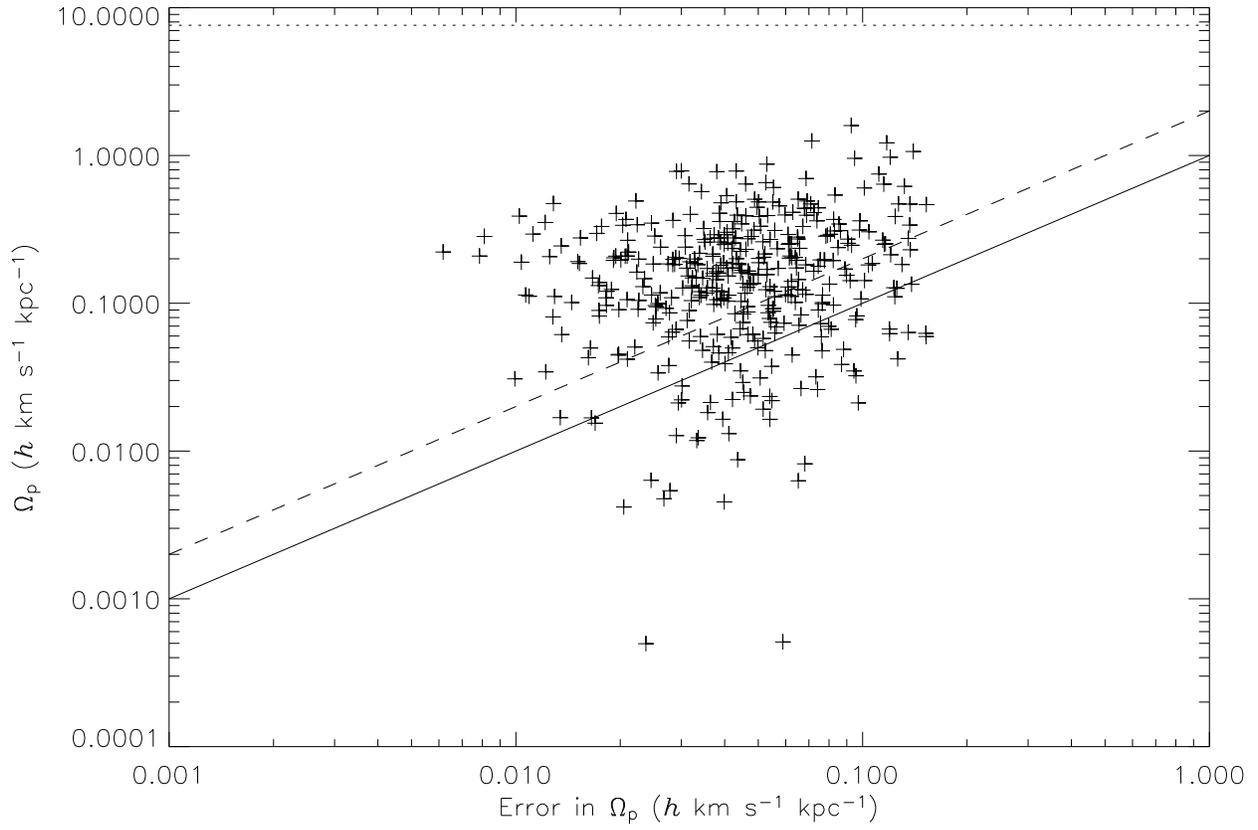}
\caption{\label{omp vs err b096}%
As in Figure~\ref{omp vs err}, but only including snapshots
b096 through b102.}
\end{figure}

The dotted line in both Figures~\ref{omp vs err} and~\ref{omp hist}
shows the Nyquist frequency of 3.8~\hkmskpc.
If the
measured distribution of pattern speeds extended up to the Nyquist frequency,
the intrinsic distribution would likely extend above the Nyquist frequency,
and the results would be affected by frequency aliasing.
However, the measured distribution does not approach
the Nyquist frequency.
Therefore, any halo whose figure rotation is aliased
would need to be wildly anomalous, with a pattern speed
many times faster than any other halo in our sample.
We consider this unlikely.
Figure~\ref{omp vs err b096} shows the pattern speeds as a function
of the error, as in
Figure~\ref{omp vs err}, except that it only uses snapshots
b096 through b102, so the maximum time between snapshots is
200~\hmyr, and the corresponding Nyquist frequency is
7.6~\hkmskpc, shown again as the dotted line. The top of the
distribution does not change between
Figures~\ref{omp vs err} and~\ref{omp vs err b096}, demonstrating
that the results are not affected by aliasing.

\begin{figure}
\plotone{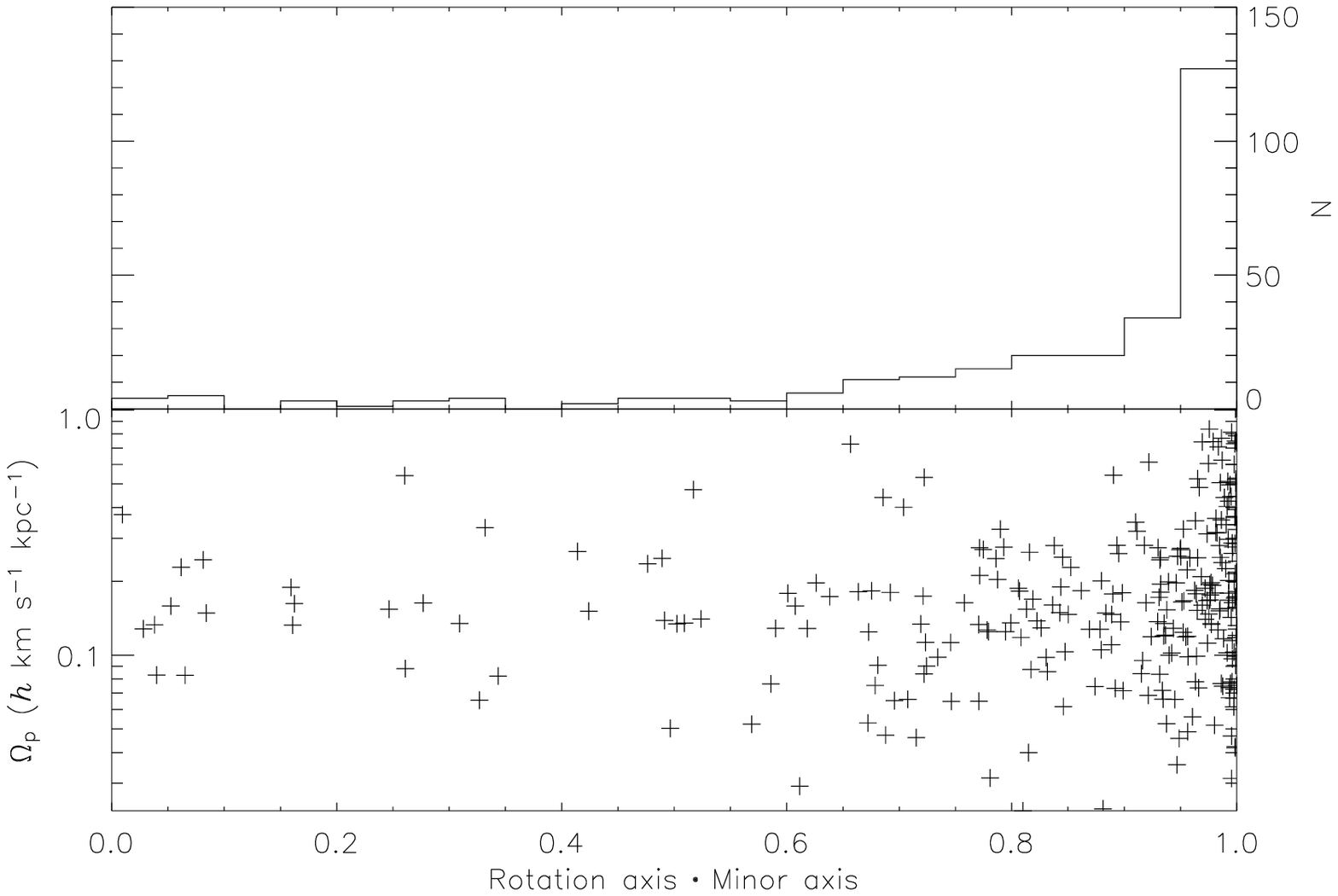}
\caption{\label{rot vs min}%
\textit{(Top)}: Histogram of the direction cosine between
the minor axis of the halo and the figure rotation axis,
for halos with $2\sigma$ detections of figure rotation.
Because the minor axis has reflection symmetry,
this is always positive.
\textit{(Bottom)}: Direction cosine as a function of
the magnitude of figure rotation.}
\end{figure}

We investigate how the figure rotation axis
relates to two other important axes. Both \citetalias{dubinski92}
and \citetalias{pfitzner99} claim that the major axis rotates
around the minor axis. The direction cosine between the
rotation axis and the minor axis is plotted both
as a function of the pattern speed and as a histogram
in Figure~\ref{rot vs min}.
We confirm that the rotation
axis coincides very strongly with the minor axis.
Due to the tendency for the halos to be prolate, the minor and
intermediate axes tend to degenerate, accounting for
some (but not all) of the misaligned tail.

\begin{figure}
\plotone{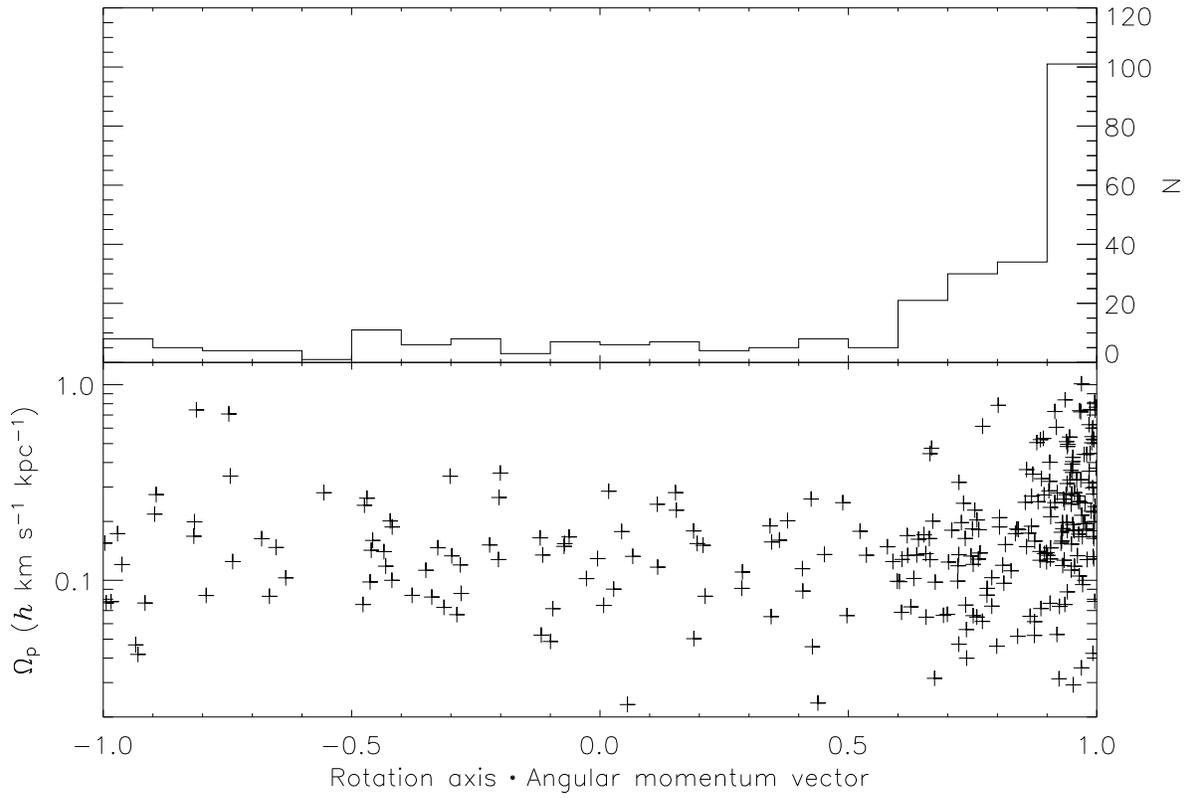}
\caption{\label{rot vs L}%
\textit{(Top)}: Histogram of the direction cosine between
the angular momentum vector of the halo and the
figure rotation axis,
for halos with $2\sigma$ detections of figure rotation.
\textit{(Bottom)}: Direction cosine as a function of
the magnitude of figure rotation.}
\end{figure}

The rotation axis is compared to the angular momentum vector of the
halo in Figure~\ref{rot vs L}. Because the angular momentum is usually
well aligned with the minor axis of halos \citep{warren-etal92}, it is no
surprise that the rotation axis is also well aligned with the angular
momentum vector.
Because the alignment between the minor axis and angular momentum of halos is
not perfect
(\citet{warren-etal92}, Bailin~\& Steinmetz, in preparation), 
some of the halos with perfectly aligned figure rotation and minor axes
have less perfect alignment between the figure rotation axis and the
angular momentum vector, seen as the bump in Figure~\ref{rot vs L}
that extends down to a direction cosine of 0.6;
the tail of the distribution extends all the way to anti-alignment.
The alignment is also plotted as a function of the pattern speed in the
lower panel of Figure~\ref{rot vs L}.
There is no trend for the halos with slow figure rotation, but all but two
of the halos with $\Omega_p > 0.4~\hkmskpc$ have figure rotation axes and
angular momentum vectors that are well aligned, with a direction cosine
of 0.65 or higher.

\begin{figure}
\plotone{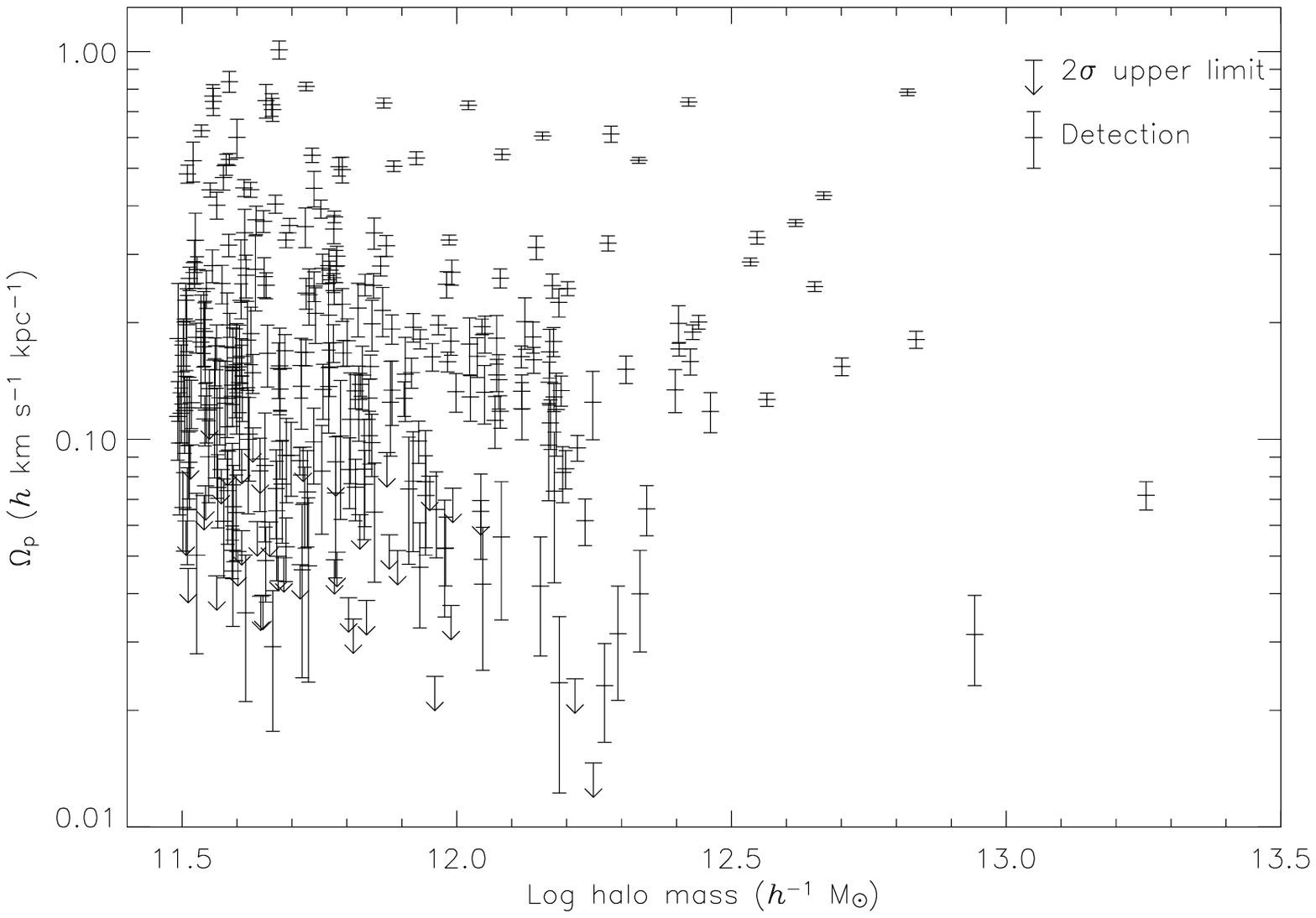}
\caption{\label{omp vs mass}%
Pattern speed of the figure rotation versus the mass of the halo.
Error bars are $1\sigma$ errors for halos with at least $2\sigma$
detections of figure rotation.
The upper limits are the halos
with a measured $\Omega_p < 2\sigma$, and are plotted at the
$2\sigma$ limit.%
}
\end{figure}

\begin{figure}
\plotone{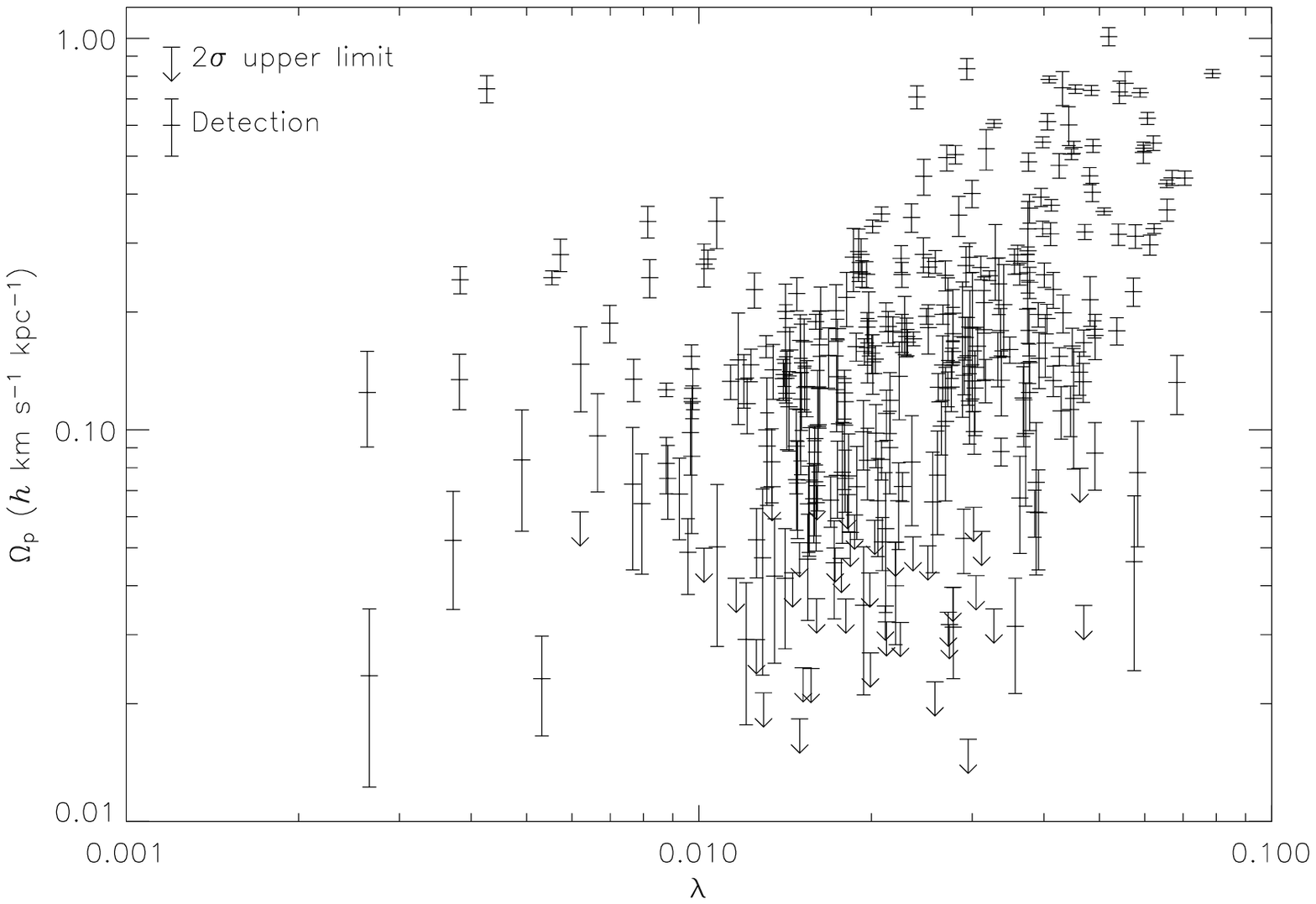}
\caption{\label{omp vs spin}%
Pattern speed of the figure rotation versus the spin parameter
of the halo. Error bars and upper limits are as in
Figure~\ref{omp vs mass}.%
}
\end{figure}

We have attempted to see if the pattern speed is correlated
with other halo properties, in particular its
mass and its angular momentum. Figure~\ref{omp vs mass} shows the
pattern speed of the figure rotation versus the halo mass. Error
bars are $1\sigma$ errors, with $2\sigma$ upper limits plotted
for halos which lie below the dashed line of Figure~\ref{omp vs err}.
There is no apparent correlation between the halo mass and its
pattern speed.

Figure~\ref{omp vs spin} shows the pattern speed versus the spin
parameter $\lambda$, where
\begin{equation}\label{peebles lambda}
\lambda \equiv \frac{ J \left| E \right|^{1/2} } { G M^{5/2} }
\end{equation}
\citep{peebles69}.
We use the computationally simpler $\lambda'$ as an estimate for $\lambda$,
where
\begin{equation}\label{lambda prime defn}
\lambda' \equiv \frac{J}{\sqrt{2} M V R}
\end{equation}
\citep{bullock-etal01}.
There is a tendency for halos with fast figure rotation to have
large spin parameters; in particular, all but one of the halos
with $\Omega_p > 0.4~\hkmskpc$ have $\lambda > 0.024$. These are
the same halos which are shown to have particularly well-aligned
rotation axes and angular momentum vectors in Figure~\ref{rot vs L}.
We have calculated the median value of $\Omega_p$ including the upper
limits for bins of width $\Delta\lambda = 0.01$. The median rises
steadily from 0.12~\hkmskpc\ for $\lambda<0.02$ to 0.44~\hkmskpc\ for
$\lambda > 0.06$.
Note that \citetalias{pfitzner99} only detected figure rotation in
halos with $\lambda>0.05$ (see his Figure~5.24).

\begin{figure}
\plotone{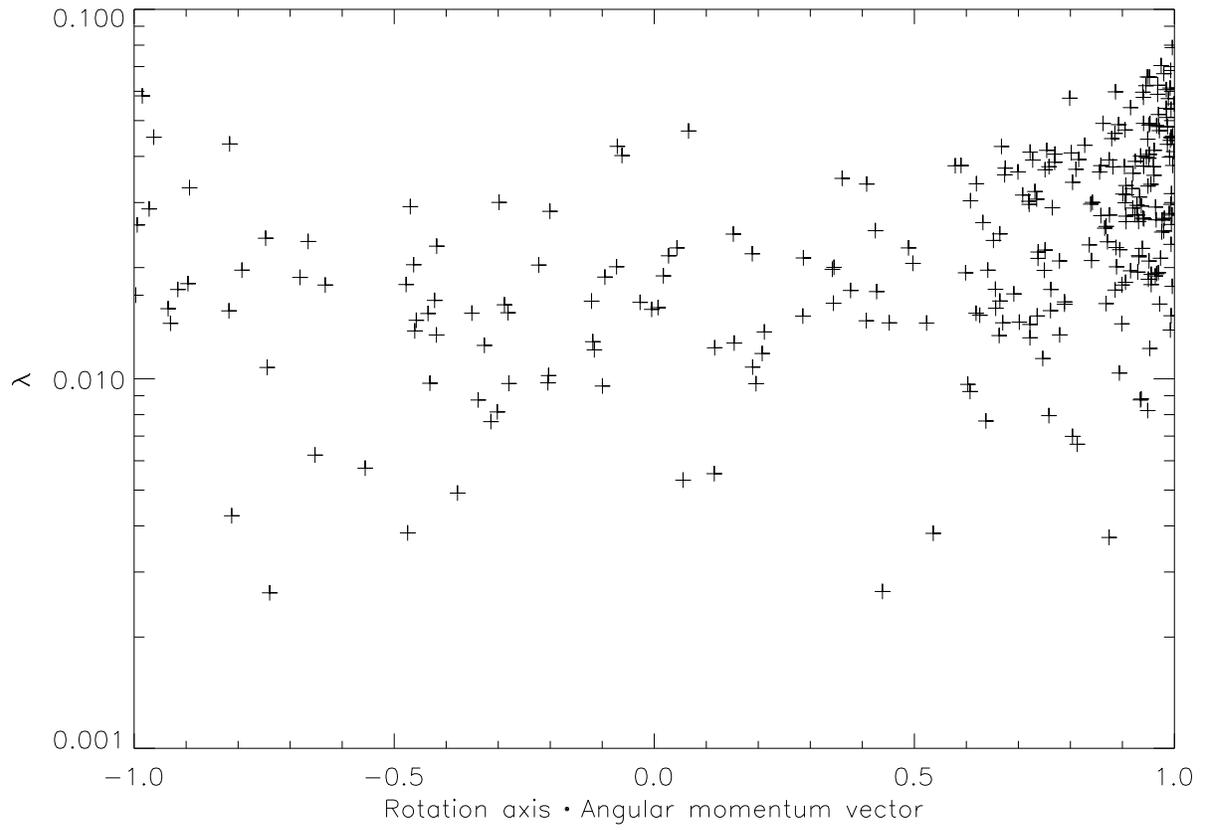}
\caption{\label{rot vs L vs lambda}%
Direction cosine between the angular momentum vector of the halo
and the figure rotation axis versus the spin parameter
$\lambda$.}
\end{figure}

The degree of alignment between the figure rotation axis and the
angular momentum vector may depend on the angular momentum content
of the halo.
Figure~\ref{rot vs L vs lambda} shows how this alignment depends
on the spin
parameter $\lambda$. There is no trend for $\lambda < 0.05$, but
the halos with $\lambda > 0.05$ show particularly good
alignment.
This is a natural consequence of the tendency for halos with rapid
figure rotation to have well aligned figure rotation and angular momentum
axes (Figure~\ref{rot vs L}), and the correlation between figure pattern
speed and spin parameter $\lambda$ (Figure~\ref{omp vs spin}).

\begin{figure}
\plotone{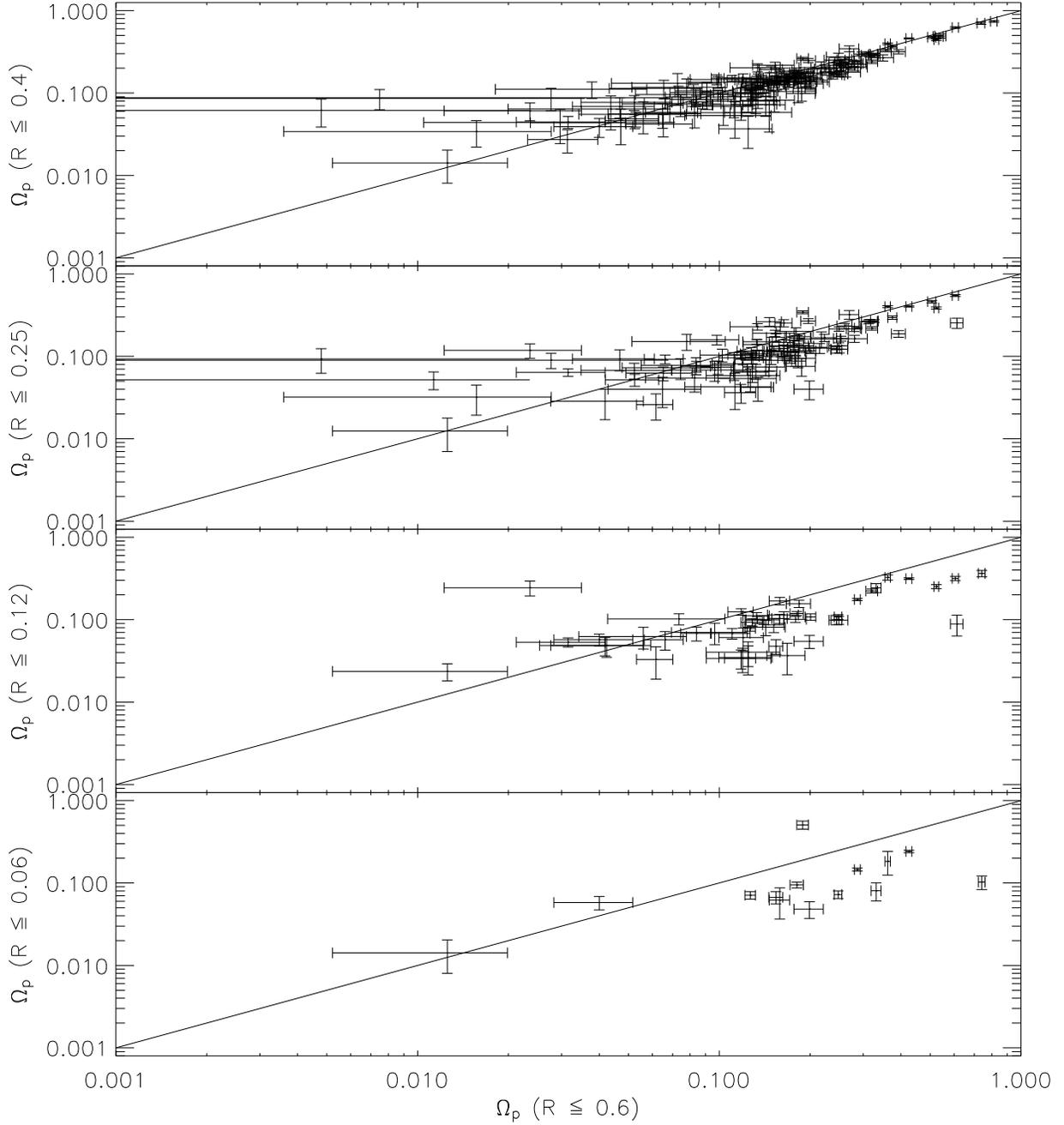}
\caption{\label{radial comparison}%
Pattern speed of figure rotation $\Omega_p$ at 0.4, 0.25, 0.12, and 0.06 of the
virial radius \rvir\ (top to bottom) as a function of the pattern speed
at $0.6~\rvir$.
Only halos where both radii in the comparison
contain at least 4000~particles, pass all of the tests of
Section~\ref{5sigma deviations}, and have $2\sigma$ detections of
figure rotation are included.
All units are \hkmskpc. The solid line corresponds
to equal pattern speeds.
}
\end{figure}

\begin{figure}
\plotone{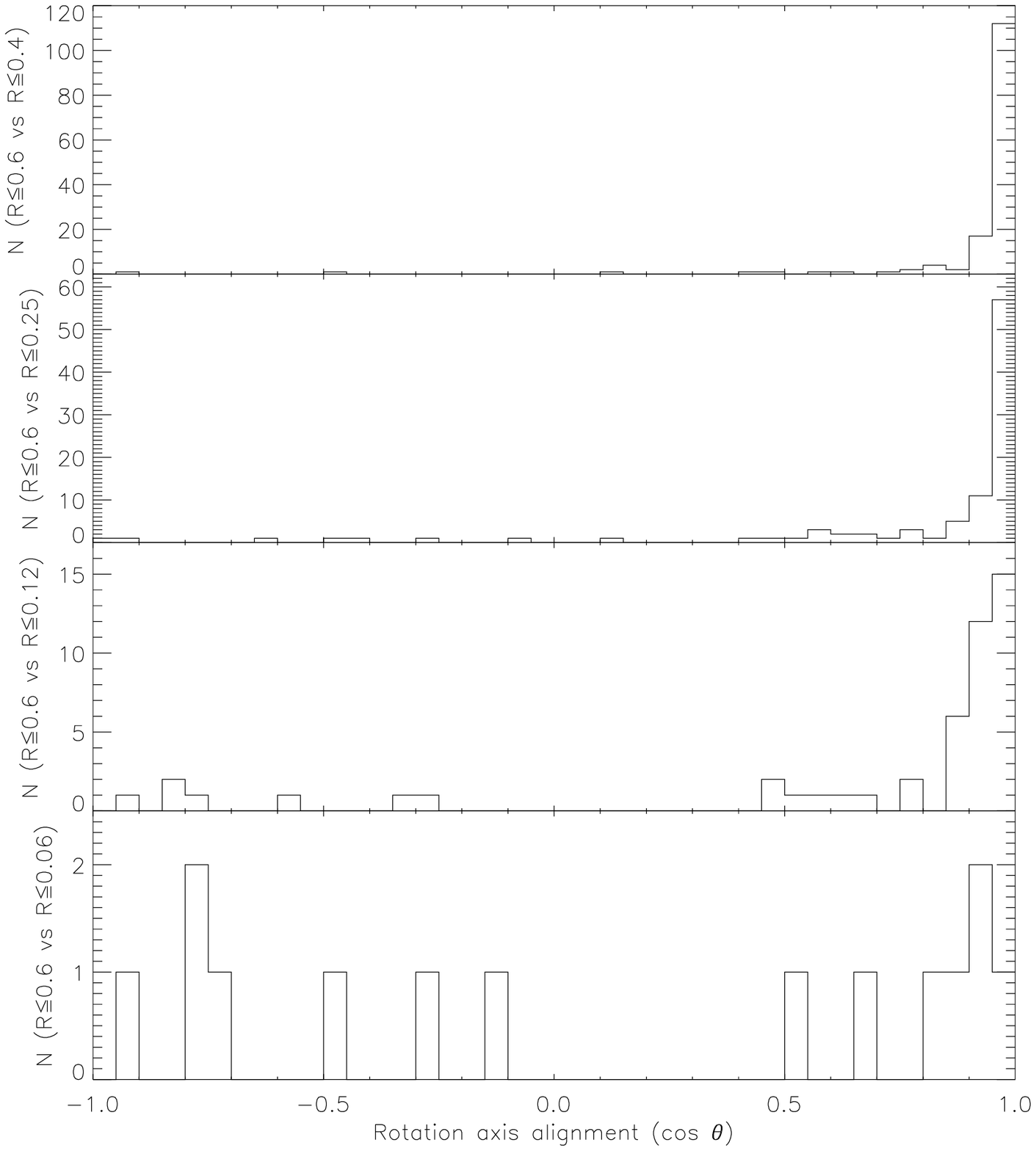}
\caption{\label{radial alignment}%
Histogram of the direction cosine of the alignment between the figure
rotation axis of the halo as a whole ($r < 0.6~\rvir$) and of just
the inner 0.4, 0.25, 0.12, and 0.06 of the virial radius
\rvir\ (top to bottom).
Only halos where both radii in
the comparison contain at least 4000~particles, pass all
of the tests of Section~\ref{5sigma deviations},
and have $2\sigma$ detections of figure rotation are included.
}
\end{figure}

Figures~\ref{radial comparison} and~\ref{radial alignment}
show how the figure rotation changes
with radius.
Figure~\ref{radial comparison} shows how the pattern speeds at different
radii are related, while Figure~\ref{radial alignment} shows the
alignment of the figure rotation axes between radii.
Each panel includes only the halos that have at least
4000~particles within the inner radius, pass all of the
tests of Section~\ref{5sigma deviations} for both radii,
and have $2\sigma$ measurements of figure rotation at both radii. Due to
the smaller number of particles in spheres of smaller radii, there
are progressively fewer halos with good measurements at smaller
radii.
The top panels show that the figure rotation in the outer regions
of the halo is very coherent. To some degree,
this is by construction; the test for substructure is equivalent
to a cut in $\Delta \Omega_p$ between adjacent radii. However,
gradual
drifts of $\Omega_p$ and changes in the figure rotation axis
with radius are still possible. The lower panels
of Figures~\ref{radial comparison} and~\ref{radial alignment}
show that this indeed happens.
In particular, while the figure rotation within 0.12~\rvir\ and
0.6~\rvir\ are strongly correlated, the pattern speeds within 0.12~\rvir\ 
are slightly smaller and the alignment
of the rotation axes is not quite as strong.
The bottom panels show that
in the innermost regions, within 0.06~\rvir, the pattern speeds are
significantly smaller than for the halo as a whole, particularly for those
halos with high pattern speeds, and more than half of the halos show
no alignment between the figure rotation axes.

We examine three possible explanations for these trends with radius.
First, it may be that the halos with high pattern speeds are still
affected by residual substructure in the outer regions.
However, the gradual decline for all halos seen as the radius shrinks suggests
that the mechanism responsible for the difference affects all halos
equally and gradually, rather than affecting a few halos at a specific
radius. 
Another piece of evidence that argues against this explanation is that
the halos with the highest measured pattern speeds
do not have preferentially high values of $f_s$; they have values
evenly spread between 0 and the cutoff of 0.05.
A second possibility is that the figure rotation, although steady
on timescales of 1~Gyr, may be fundamentally a transitory feature caused
by a tidal encounter or the most recent major merger.
The inner region of the halo has a shorter dynamical time, and therefore
the effects of such a disturbance will be erased faster
in the inner regions than the outer regions.
This is consistent with the gradual decrease in pattern speed
with radius and the decrease in alignment.
However, the halo of \citetalias{bureau-etal99}
has fast figure rotation (faster than any of our halos), and yet
shows steady figure rotation at all radii for 5~Gyr.
We propose instead that the effects of force softening are becoming important
at the smaller radii.
The radius of the innermost sphere for the halos plotted in
the bottom panels of Figures~\ref{radial comparison} and~\ref{radial alignment}
range from 3--5 force softening lengths, where the effects of the gravitational
softening can still be important \citep{power-etal03}. The weaker
gravitational force results in a more spherical potential, consistent
with the weaker figure rotation and lack of alignment.

\begin{figure}
\plottwo{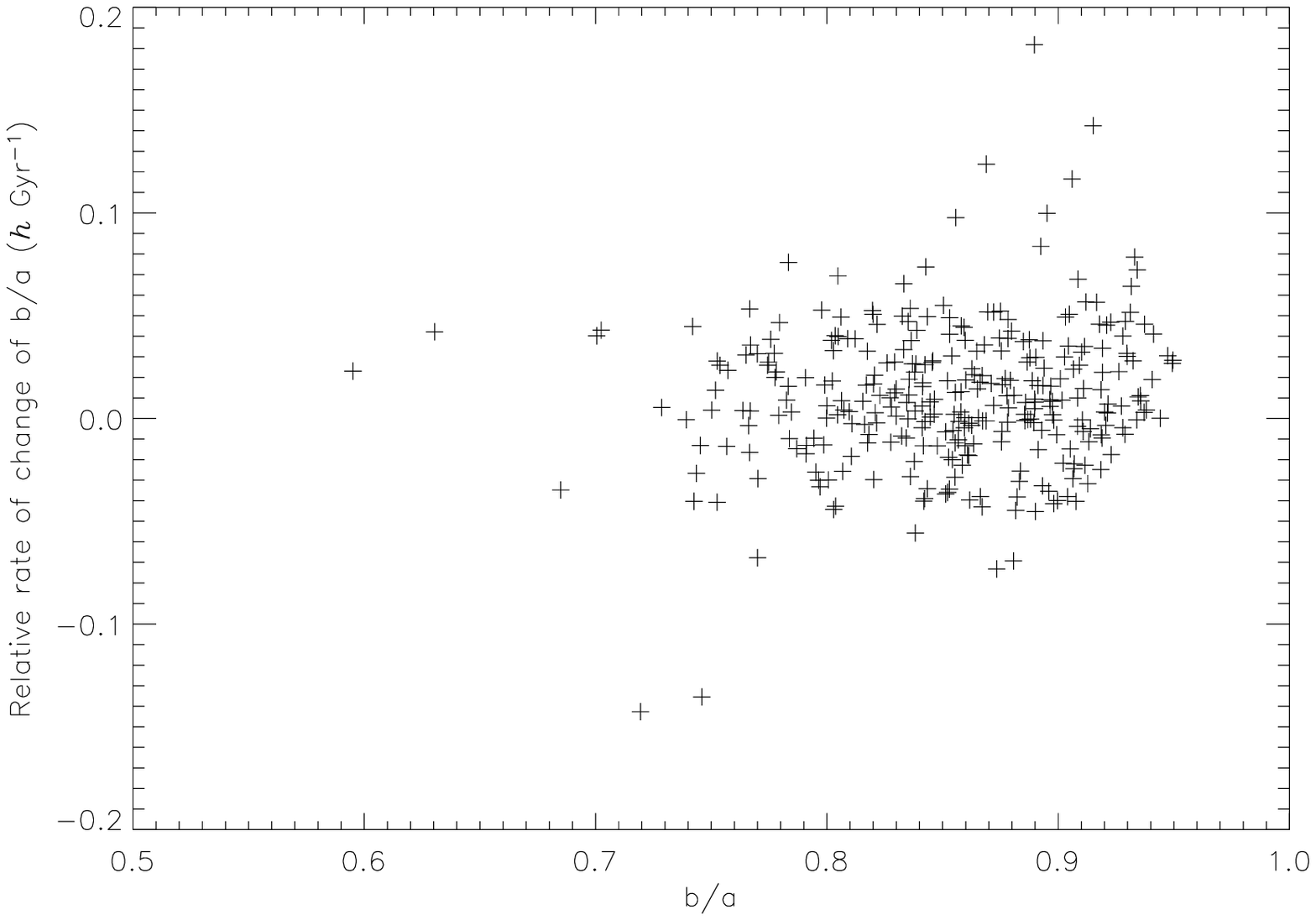}{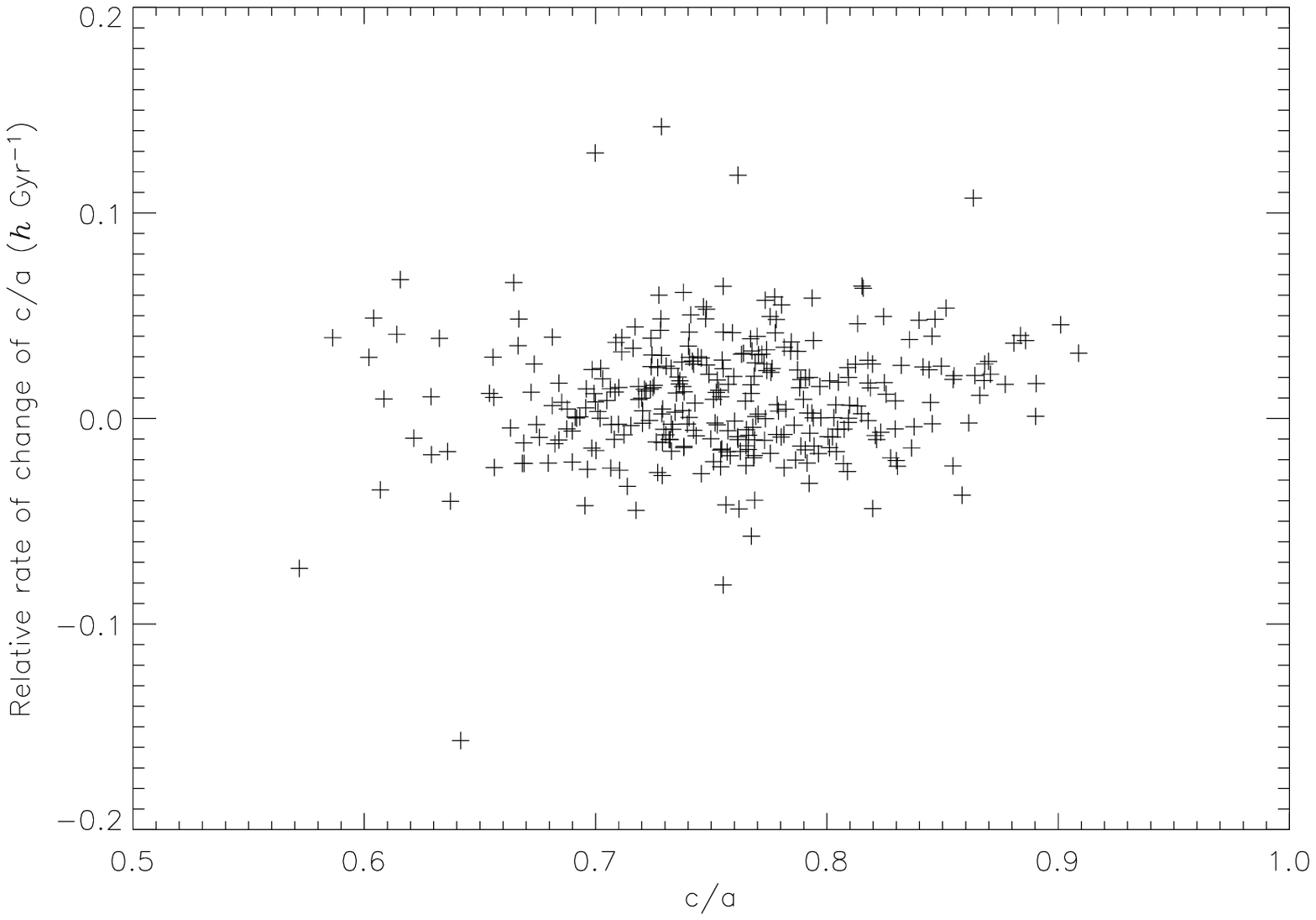}
\caption{\label{axis ratio change}%
Fractional
rate of change of the $b/a$ \textit{(left)} and $c/a$ \textit{(right)}
axis ratio of the halos in the sample,
as a function of the $z=0$ axis ratio.}
\end{figure}

We have calculated the rate of change of the $b/a$ and $c/a$ axis ratios
over the 5 snapshots using linear regression. The evolution of the axis
ratios with time is linear for almost all halos. Figure~\ref{axis ratio change}
shows the fractional rate of change, $(\dot{b/a}) / (b/a)$ and
$(\dot{c/a}) / (c/a)$, of the axis
ratios
 as a function of the value of the axis ratio in the final snapshot.
The median and standard deviation of the distribution of 
$(\dot{b/a}) / (b/a)$
are $0.0089~h~\mathrm{Gyr^{-1}}$ and $0.0349~h~\mathrm{Gyr^{-1}}$ respectively.
For $c/a$, they are $0.0093~h~\mathrm{Gyr^{-1}}$ and
$0.0297~h~\mathrm{Gyr^{-1}}$ respectively. Therefore, there is a
weak tendency for undisturbed halos to become more spherical
with time. Most halos require several Gyr before their flattening changes
significantly;
there are, however, a few outliers
with quite significant changes in their axis ratios.
Figure~\ref{axis ratio change} demonstrates that there is no trend of
$(\dot{b/a}) / (b/a)$ or $(\dot{c/a}) / (c/a)$ with the value of the axis ratio
except for the outliers with very high (low) values of $\dot{b/a}$,
which could not have such high (low) rates of change if the values
of $b/a$ were not very high (low) in the final snapshot.
We find no trend with any other halo
property such as mass, spin parameter, pattern speed, substructure fraction,
or alignment of the figure rotation axis with the angular momentum vector or
minor axis.

\section{Conclusions}\label{conclusions section}

We have detected rotation
of the orientation of the major axis in most undisturbed halos of
a \lcdm\ cosmological simulation. The axis around
which the figure rotates
is very well aligned with the minor axis in most cases. It
is also usually well aligned with the angular momentum vector.
The distribution of pattern speeds is well fit by a log
normal distribution,
\begin{equation}
P(\Omega_p) = \frac{1}{\Omega_p \sigma \sqrt{2\pi}}
  \exp\left(-\frac{\ln^2(\Omega_p/\Omega_{p_0})} {2\sigma^2} \right),
\end{equation}
with $\Omega_{p_0}=0.148~\hkmskpc$ and $\sigma=0.83$.

The pattern speed $\Omega_p$ is correlated with spin parameter $\lambda$.
The median pattern speed rises from 0.12~\hkmskpc\ for halos
with $\lambda < 0.02$
to 0.44~\hkmskpc\ for halos with $\lambda > 0.06$,
with a spread of almost an order of magnitude around this median
at a given value of $\lambda$.
The 11\%\ of halos in the sample
with the highest pattern speeds, $\Omega_p > 0.4~\hkmskpc$,
not only have large spin parameters, but also show particularly strong
alignment between their figure rotation axes and their angular
momentum vectors.
There is no obvious correlation of the figure rotation properties with mass.
The pattern speed and figure rotation axis is coherent in the outer regions
of the halo.
Within 0.12~\rvir, the pattern speed drops, particularly for those
halos with fast figure rotation, and the
internal alignment of the figure rotation axis deteriorates.
This is probably an artifact of the numerical force softening.

\citetalias{bureau-etal99} hypothesized that
the spiral structure in \ntnif\ is due to figure rotation of
a triaxial halo.
The required pattern speed of $7\pm1~\kmskpc$ \citepalias{mb03}
is much higher than the pattern speeds seen in the simulated halos,
and is estimated to have a probability of $5 \times 10^{-7}$. We therefore
conclude that the figure rotation of
undisturbed \lcdm\ halos is not able to produce this spiral structure.
Halos with
large values of $\lambda$ tend to have more substructure \citep{be87},
so there is a deficiency of halos with very high $\lambda$ in our sample.
Because $\Omega_p$ correlates with $\lambda$, we cannot exclude the
possibility that there exist halos with very high $\lambda$ whose
figures rotate sufficiently quickly.
However, halos with such high $\lambda$ are themselves
very rare \citepalias{mb03}, and if such halos fall out of our sample
due to the presence of strong substructure, the effects of the substructure
on the gas disk of \ntnif\ would be of more concern than the slow
rotation of the halo figure, a possibility \citetalias{bureau-etal99}
rule out due to the lack of any plausible companion in the vicinity.

More generally, \citet{bf02} found very weak if any enhancement of spiral
structure in disk simulations with triaxial figures rotating at
0.77~\kmskpc, a value similar to the highest pattern speed seen
in our sample. Therefore, it is unlikely that triaxial figure rotation
can be detected by looking for spiral structure in extended gas disks.

We have found that the axis ratios of undisturbed halos tend to become
more spherical with time, with median fractional increases in the
$b/a$ and $c/a$ axis ratios of $\approx 0.009~h~\mathrm{Gyr^{-1}}$.
The distributions of $(\dot{b/a})/(b/a)$ and $(\dot{c/a})/(c/a)$
are relatively wide,
with standard deviations of $\approx 0.03~h~\mathrm{Gyr^{-1}}$.
A few outliers have axis ratios that change quite significantly
over the span of 1~Gyr. The rate of change of the axis ratios
is not correlated with any other halo property.

\acknowledgements
This work has been supported by grants from the U.S. National Aeronautics and
Space Administration (NAG 5-10827), the David and Lucile Packard Foundation, and by the 
Bundesministerium f\"ur Bildung und Forschung (FKZ 05EA2BA1/8).
We thank Volker Springel for providing us with an advance version
of GADGET2.
JB thanks Chris Power and Brad Gibson for useful discussions. We furthermore thank
Ken Freeman and Martin Bureau for their repeated exhortations to
pursue these calculations.

\bibliography{ms}

\end{document}